\documentclass{aa}  
\usepackage{graphicx}
\usepackage{xcolor}
\usepackage{txfonts}
\usepackage{pifont}
\usepackage{keyval}
\usepackage[colorlinks=true,linkcolor=blue,filecolor=blue,citecolor = blue,urlcolor = blue]{hyperref}
\usepackage{hyperref}
\usepackage{amsmath}
\usepackage{makecell}
\usepackage{subcaption}

\defcitealias{Naidu+2024_alt}{Naidu \& Matthee et al. 2024} 
\begin{document}

   \title{The slope and scatter of the star forming main sequence at $z\sim5$ : reconciling observations with simulations}
   \titlerunning{The slope and scatter of the star forming main sequence at $z\sim5$}

   \author{Claudia~Di~Cesare\inst{\ref{inst:1}}\fnmsep\thanks{email: claudia.dicesare@ista.ac.at}
   \and Jorryt~Matthee\inst{\ref{inst:1}}
   \and Rohan~P.~Naidu \inst{\ref{inst:2}}
   \and Alberto~Torralba\inst{\ref{inst:1}}
   \and Gauri~Kotiwale\inst{\ref{inst:1}}
   \and Ivan~G.~Kramarenko\inst{\ref{inst:1}}
   \and Jeremy~Blaizot\inst{\ref{inst:3}}
   \and Joakim~Rosdahl\inst{\ref{inst:3}}
   \and Joel~Leja\inst{\ref{inst:4},\ref{inst:5}}
   \and Edoardo~Iani\inst{\ref{inst:1}}
   \and Angela~Adamo\inst{\ref{inst:6}}
   \and Alba~Covelo-Paz\inst{\ref{inst:11}}
   \and Lukas~J.~Furtak \inst{\ref{inst:7}}
   \and Kasper~E.~Heintz\inst{\ref{inst:8},\ref{inst:9},\ref{inst:10}}
   \and Sara~Mascia\inst{\ref{inst:1}}
   \and Benjamín~Navarrete\inst{\ref{inst:1}}
   \and Pascal~A.~Oesch\inst{\ref{inst:11},\ref{inst:12},\ref{inst:13}}
   \and Michael~Romano\inst{\ref{inst:14}, \ref{inst:15}}
   \and Irene~Shivaei\inst{\ref{inst:16}}
   \and Sandro~Tacchella\inst{\ref{inst:17}, \ref{inst:18}}
          }

   \institute{
   Institute of Science and Technology Austria (ISTA), Am Campus 1, 3400 Klosterneuburg, Austria \label{inst:1} 
   \and MIT Kavli Institute for Astrophysics and Space Research, Massachusetts Institute of Technology, Cambridge, MA 02139, USA \label{inst:2}
   \and Universite Claude Bernard Lyon 1, CRAL UMR5574, ENS de Lyon, CNRS, Villeurbanne, F-69622, France \label{inst:3}
   \and Institute for Computational \& Data Sciences, The Pennsylvania State University, University Park, PA 16802, USA \label{inst:4}
   \and Institute for Gravitation and the Cosmos, The Pennsylvania State University, University Park, PA 16802, USA \label{inst:5}
   \and Department of Astronomy, Oskar Klein Centre, Stockholm University, AlbaNova University Center, SE-106 91 Stockholm, Sweden \label{inst:6}
   \and Physics Department, Ben-Gurion University of the Negev, P.O. Box 653, Beer-Sheva 8410501, Israel \label{inst:7}
   \and Niels Bohr Institute, University of Copenhagen, Jagtvej 128, 2200 Copenhagen N, Denmark \label{inst:8}
   \and Department of Astronomy, University of Geneva, Chemin Pegasi 51, 1290 Versoix, Switzerland \label{inst:9}
   \and Laboratory of Astrophysics, \'Ecole Polytechnique F\'ed\'erale de Lausanne (EPFL), Observatoire de Sauverny, 1290 Versoix, Switzerland \label{inst:10}
   \and Department of Astronomy, University of Geneva, Chemin Pegasi 51, 1290 Versoix, Switzerland \label{inst:11}
   \and Cosmic Dawn Center (DAWN), Copenhagen, Denmark \label{inst:12}
   \and Niels Bohr Institute, University of Copenhagen, Jagtvej 128, K{\o}benhavn N, DK-2200, Denmark \label{inst:13}
   \and Max-Planck-Institut f\"ur Radioastronomie, Auf dem H\"ugel 69, 53121 Bonn, Germany\label{inst:14} 
   \and INAF - Osservatorio Astronomico di PAdova, Vicolo dell'Osservatorio 5, I-35122 Padova, Italy\label{inst:15} 
    \and Centro de Astrobiolog\'ia (CAB), CSIC-INTA, Carretera de Ajalvir km 4, Torrej\'on de Ardoz, E-28850, Madrid, Spain \label{inst:16}
   \and Kavli Institute for Cosmology, University of Cambridge, Madingley Road, Cambridge, CB3 0HA, UK \label{inst:17}
   \and Cavendish Laboratory, University of Cambridge, JJ Thomson Avenue, Cambridge, CB3 0HE, UK \label{inst:18}
             }

   \date{}

   \abstract{

   Galaxies exhibit a tight correlation between their star-formation rate and stellar mass over a wide redshift range known as the star-forming main sequence (SFMS). With JWST, we can now investigate the SFMS at high redshifts down to masses of $\sim10^6$ M$_{\odot}$, using sensitive star-formation rate tracers such as H$\alpha$ emission -- which allow us to probe the variability in star formation histories. We present inferences of the SFMS based on 316 H$\alpha$-selected galaxies at $z\sim4-5$ with log($\rm M_\star/M_\odot) = 6.4 -10.6$. These galaxies were identified behind the Abell 2744 lensing cluster with NIRCam grism spectroscopy from the “All the Little Things” (ALT) survey. At face value, our data suggest a shallow slope of the SFMS (SFR $\propto \mathrm M_\star ^\alpha$, with $\alpha=0.45$). After correcting for the H$\alpha$-flux limited nature of our survey using a Bayesian framework, the slope steepens to $\alpha$ = 0.59$^{+0.10}_{-0.09}$, whereas current data on their own are inconclusive on the mass dependence of the scatter. These slopes differ significantly from the slope of $\approx1$ expected from the observed evolution of the galaxy stellar mass function and from simulations. When fixing the slope to $\alpha=1$, we find evidence for a decreasing intrinsic scatter with stellar mass (from $\approx$ 0.5 dex at M$_\star=10^8$ M$_\odot$ to 0.4 dex at M$_\star= 10^{10}$ M$_\odot$). This tension might be explained by a (combination of) luminosity-dependent SFR(H$\alpha$) calibration, a population of (mini)-quenched low-mass galaxies, or underestimated dust attenuation in high-mass galaxies. Future deep observations across facilities can quantify these processes, enabling better insights into the variability of star formation histories. 
}


   \maketitle
%
\section{Introduction}
Star formation is a key process in galaxies, shaping their physical properties over cosmic time. The star-formation rate (SFR) has an impact on the chemical enrichment of their interstellar and circumgalactic medium \citep[ISM/CGM;][]{Oppenheimer+2010,Ginolfi+2020}, the dynamical processes within galaxies \citep{Hung+2019, Danhaive+2025}, the reionization of the Universe \citep[e.g.,][]{Simmonds+2024} and it controls the rate of galaxy assembly over cosmic time \citep[e.g.,][]{Madau+Dickinson2014}. 

The SFR of galaxies is probed using various diagnostics: UV-continuum observations, nebular emission lines, far-infrared continuum observations \citep{Kennicutt1998a,KennicuttEvans2012}, mid-infrared features (i.e. polycyclic aromatic hydrocarbon, \citealt{Shipley+2016}) and radio emission \citep[e.g.][]{Duncan+2020}. In this work, we focus on nebular emission lines, which arise from H\,{\sc ii} regions around young, massive stars ($\rm M_s \gtrsim 10 \; M_\odot$, main-sequence lifetime $\lesssim 10$ Myr). The recombination of ionized gas surrounding these stars produces hydrogen emission lines, such as Balmer lines, which can be used as SFR diagnostics, as their flux is proportional to the incident ionizing continuum. Since only massive stars contribute significantly to the Lyman continuum (LyC) luminosity, nebular lines trace the SFR on short timescales.

In the context of galaxy evolution, a key relation to investigate is the one between stellar mass ($\rm M_\star$) and SFR commonly referred to as the “main sequence of star-forming galaxies” \citep[e.g.][]{Noeske+2007}, which we will refer to as star-forming main sequence (SFMS) or SFR-M$_\star$ throughout the paper. The normalization of the SFMS together with its slope and scatter encapsulate information on the mechanisms and efficiency of gas conversion into stars \citep[e.g.,][]{Speagle+2014,Sparre+2015,Iyer+2018,Clarke+2024}. The existence of a tight SFR-M$_\star$ relation suggests that star formation in galaxies is self-regulated by the interplay of gas accretion, star formation and feedback processes \citep{Tacchella+2016a}. As such, the distribution of SFR at fixed stellar mass and redshift is shaped by short-timescales variations driven by feedback duty-cycle \citep[e.g.,][]{Peng+2010} and by long-term processes linked to variations in halo accretion rates \citep[e.g.,][]{ Abramson+2016,Matthee+2019, Tacchella+2020_sfhII}.

The tight relation between SFR and M$_\star$ is observed from $z \sim 0$ to at least $z \sim 6$, and its redshift evolution has been widely studied using both observations \citep[e.g.,][]{Whitaker+2012_sfms, Speagle+2014, Popesso+2023} and simulations \citep[e.g.,][]{Pillepich+2019_tng50, DiCesare+2023, Katz+2023_sphinx, McClymont+2025}. The precise functional form of the SFMS remains uncertain: while some studies characterize it as a simple power law across all stellar masses \citep[$\mathrm{log(M_\star/M_\odot) >8}$,][]{Speagle+2014, Pearson+2018_ms}, others report evidence for a high-mass turnover \citep{Lee+2015,Tomczak+2016_msbending, Popesso+2019_ms,Popesso+2023}. The turnover mass is above $\mathrm{M^{turn}_\star = 3 \times 10^{10} \; M_\odot}$ for $z > 4$ \citep{Popesso+2023}, possibly associated with bulge formation. Despite discrepancies at the high mass end, there is broad consensus that, over the stellar mass range $\sim 10^8$–$10^{11} ~ \mathrm{M_\odot}$, the SFR–$\mathrm{M_\star}$ relation is well-described by a power-law.

The SFMS slope is closely related to the growth of the galaxy stellar mass function (SMF) since at each redshift, mass is added to the mass function by star formation \citep{Peng+2010, Leja+2015}. \citet{Leja+2015} analyzed the connection between the observed SFMS (SFR $\propto \rm M_\star ^\alpha$) and the observed evolution of the stellar mass function at $0.2 < z < 2.5$ showing that the slope cannot have values $< 0.9$ at all masses and redshifts, as this would result in a much higher number density of low-mass galaxies than observed. From theory, the normalization of the SFMS is expected to evolve rapidly with redshift, driven by the increase in baryon accretion rates at earlier times, following $(1+z)^\gamma$ with $\gamma \approx 2.5$ \citep{Dave+2011,Dayal+2013, Dekel+2013, Sparre+2015,Tacchella+2016a}. Observations from the REBELS survey \citep{Bouwens+2022_rebels} pointed towards a slower evolution of the normalization with redshift \citep[$\gamma \sim 1.6$;][]{Topping+2022}. However, more recent estimates from both simulations and observations found a stronger redshift evolution, with $\gamma \sim 2.6$ \citep{McClymont+2025} and $\gamma = 2.3$ \citep{Simmonds+2025}, respectively. There is still no clear consensus in the literature on the value of $\gamma$ \citep{Speagle+2014, Whitaker+2014, Ilbert+2015,Tasca+2015,Khusanova+2021,Topping+2022,Leja+2022, McClymont+2025}. 

Finally, the (intrinsic) scatter around the SFMS is directly related to the fluctuations in the star formation history (SFH) of individual galaxies (e.g. bursty SF), which encodes information on the baryon-cycle, galaxy-galaxy mergers, strength of stellar and black hole feedback, and dark-matter accretion histories \citep{Iyer+2020_sfh,Tacchella+2020_sfhII}\footnote{In observations, the total scatter of the SFMS consists of both the measurement uncertainty and the intrinsic scatter and its estimate is complicated by sample selection issues.}. Theoretical work indicates that star formation is bursty in low-mass galaxies as a consequence of feedback which expels gas from the ISM \citep{Faucher-Giguere2018}, and across all masses when going to the high redshift Universe, because of variable gas accretion and short equilibrium timescales \citep[e.g.,][]{FurlanettoMirocha2022, Pallottini+2023, Bhagwat+2024_feedback}. Observationally, the short-term variability (i.e. burstiness) of star formation can be inferred by comparing shorter to longer timescales SFR tracers, such as the flux ratio between the H$\alpha$ emission line and the UV-continuum \citep[e.g.,][]{Emami+2019,Faisst+2019,CaplarTacchella2019}. 

Prior to JWST, SFR estimates at $z > 2.5$ relied primarily on rest-frame far-infrared (IR) observations \citep{Zavala+2021} or rest-frame UV continuum measurements, which are highly affected by dust attenuation \citep{Bouwens+2015}. However, see \citet{Rinaldi+2022_sbms} where the authors studied the SFMS of a Lyman-alpha selected sample where they expect dust to play a minor role. Other SFR estimators commonly used at low redshift, such as optical nebular emission lines, were inaccessible for HST and ground-based spectroscopy at $z \geq 2.5$. Only \textit{Spitzer} broadband photometry enabled coarse estimates of H$\alpha$ and other emission lines \citep[e.g.,][]{Caputi+2017_starbusrtMS,Faisst+2019,Bollo+2023}. Over the past two years, JWST allowed for the investigation of the H$\alpha$ emission line at $z > 2.5$ \citep[e.g.][]{Matharu+2024, Pirie+2025}. H$\alpha$ is less affected by dust than the UV continuum, 
and the detection of fainter galaxies extending the accessible mass range to much lower values ($\rm log(M_\star/M_\odot) < 8$). Recently, several JWST studies using nebular diagnostics revealed the presence of bursty star formation histories in high-$z$ galaxies \citep[e.g.,][]{Cole+2025_ceers, Dressler+2023_glassI, Dressler+2024_glassII, Caputi+2024, Clarke+2024, Ciesla+2024, Endsley+2024, Navarro-Carrera2024_burstiness, Perry+2025, Clarke+2025}, which broadens the scatter in the SFR-M$_\star$ relation.

In this work, we focus on the redshift range $z = 4-5 $, when galaxies rapidly assemble their masses, and investigate the relation between star formation and stellar mass over a wide mass range from $\rm \sim 10^{6-10} \; M_\odot$. The investigation of such low-mass galaxies is made possible by observing a lensing cluster field, where gravitational lensing from the galaxy cluster Abell 2744 provides additional magnification ($\mu$) combined with deep NIRCam grism coverage in the F356W filter. In particular, we make use of the H$\alpha$ emission line at $z = 4-5 $, from the JWST “All the Little Things” survey \citepalias[ALT;][]{Naidu+2024_alt}, the average $\mu$ for all H$\alpha$ emitters in ALT is 2.64. The paper is organized as follows. Section \ref{sec:data} presents the JWST dataset used in this work, Sect. \ref{sec:methods} presents details on the SED fitting technique, SFR estimates, and dust attenuation. In Sect. \ref{sec:bayesian_model}, we describe the Bayesian framework used to fit the observational data accounting for their biased nature. We present our results in Sect. \ref{sec:results} and discuss them, together with their implications, in Sect. \ref{sec:discussion}. A summary of our work is presented in Sect. \ref{sec:summary}.

Throughout this work, we assume a $\Lambda$CDM cosmology with parameters H$_0$ = 67.7 $\rm km \; s^{-1} \; Mpc ^{-1}$, $\rm \Omega _m = 0.31 $ and $\rm \Omega _\Lambda = 0.69 $ consistent with \citep{Planck+2020}, and we adopt a \citet{Chabrier2003} initial mass function (IMF). Magnitudes are listed in the AB system \citep{Oke+Gunn1983}. 

\section{Data}
\label{sec:data}
\subsection{ALT sample}
The main objective of this paper is to investigate the star formation of a sample of H$\alpha$ emitters (HAEs). To achieve this goal, we used observations from the Cycle 2 JWST/NIRCam Wide Field Slitless Spectroscopic (WFSS) ALT survey \citepalias[][]{Naidu+2024_alt}. The ALT survey uses NIRCam grism spectroscopy in the F356W filter\footnote{The F356W filter is where the NIRCam grism is most sensitive and has the largest fraction of the field of view with complete wavelength coverage.} ($\lambda = 3.15 - 3.96 \: \mu$m) combined with deep NIRCam imaging in all broad and medium-band filters, yielding an emission line selected sample of $\sim 1600$ galaxies with spectroscopic redshifts in the range $z \sim 0.3 - 8.5$. The survey covers an area of $\approx30$ arcmin$^2$ around the lensing cluster Abell2744 \citep[A2744;][]{Abell+1989, Merten+2011_a2744} at $z = 0.308$. Due to the magnification power of the lensing cluster and the spectral information from the grism, ALT is the optimal survey to probe the star formation rate in low-mass galaxies (down to masses $\rm M_\star \approx 6 \times 10^{6} \; M_\odot$). 

A2744 has been extensively studied in the past using various telescopes that probe different parts of the spectrum, leading to a huge amount of ancillary data around the cluster \citepalias[see][for an overview of the legacy data on this field]{Naidu+2024_alt}. Particularly relevant for this work are the JWST observations of the A2744 field from the Ultradeep NIRSpec and NIRCam Observations before the Epoch of Reionization (UNCOVER; \citealt{Bezanson+2024_uncover}) and MegaScience \citep{Suess+2024_megascience} surveys. UNCOVER and MegaScience provided us with deep NIRCam imaging of the region around A2744 in 9 broad-band and 11 medium-band filters. The photometric information from these surveys is used to tightly constrain the SED of galaxies in our sample \citepalias[see Fig. 4 of][]{Naidu+2024_alt}.

The ALT photometric catalog has been generated using all publicly available NIRCam imaging data and following the methods described in \citet{Weibel+2024}, see \citetalias[][]{Naidu+2024_alt} for detailed description. In particular, we run a median filter with a box size of 4.04"$\times$4.04" on all the images \citep[see for example][]{Bouwens+2017} and a sigma-clipping to reduce over-subtraction around stars. In the median filter-subtracted images, most of the light coming from bright foreground objects is removed to optimize the ability to detect lensed galaxies that are close to bright foreground objects. \texttt{SourceExtractor} \citep{BertinArnouts1996} has been run in dual mode with an inverse-variance weighted stack of the median subtracted images in all available long-wavelength channel (LW) filters as the detection image. All images are matched to the point spread function (PSF) resolution in the F444W filter prior to the flux extraction, which has been performed with a circular aperture of radius 0.16". To secure the detection of faint sources, whose light can be contaminated by foreground objects despite median filter subtraction, we adopted a low ($10^{-4}$) deblending parameter. Although this choice may lead to the fragmentation of clumpy galaxies into multiple components with separate entries in the catalog (see discussion in Sect. \ref{sec:Ha_emitters}), it is necessary to avoid losing high-redshift galaxies close to foreground systems.
Photometric depths range from $27.75-29.82$ magnitudes at 5$\sigma$ in 0.16" circular apertures \citepalias[see Table 1 of][ for details]{Naidu+2024_alt}. The lensing model adopted to correct the observables for magnification is an updated version of the one by \citet{Furtak+2023} described in \citet{Price+2025}.

The identification of galaxies in the NIRCam grism data has been performed using two independent and complementary approaches: \texttt{grizli} \footnote{\href{}{https://github.com/gbrammer/grizli}} and \textit{Allegro} (Kramarenko et al., in prep.). For both approaches, to facilitate the search for emission lines, a median filter (71 pixels wide with a central hole of 10 pixels) has been run on the grism images to remove continuum \footnote{Note that this removes both continuum from contaminators as well as from sources themselves.}. The key difference between \texttt{grizli} and \textit{Allegro} is that the former approach searches for spectroscopic-$z$ solutions around photometric-$z$ priors in optimally extracted 1D spectra, while the latter detects emission lines in 2D spectra and identifies fiducial lines and redshifts. We visually inspected galaxies with conflicting redshift solutions from \texttt{grizli} and \textit{Allegro}. The full ALT line-emitter catalog was created combining both approaches and implementing insights obtained through one method to inform the other \citepalias[more details can be found in ][]{Naidu+2024_alt}. The ALT DR1 catalog with source coordinates and spectroscopic redshifts is publicly available\footnote{\href{}{https://zenodo.org/records/13871850}}.
Note that the sensitivity of grism data depends on the wavelength and the position of the source in the field of view. For $\lambda = 3.15-3.95 \; \mu$m the typical 5$\sigma$ sensitivity ranges from $6-20 \times 10^{-19}$ erg s$^{-1}$ cm$^{-2}$. Thereby, the emission-line detection threshold varies with redshift as well. 

\begin{figure*}
   \centering
   \includegraphics[scale=0.4]{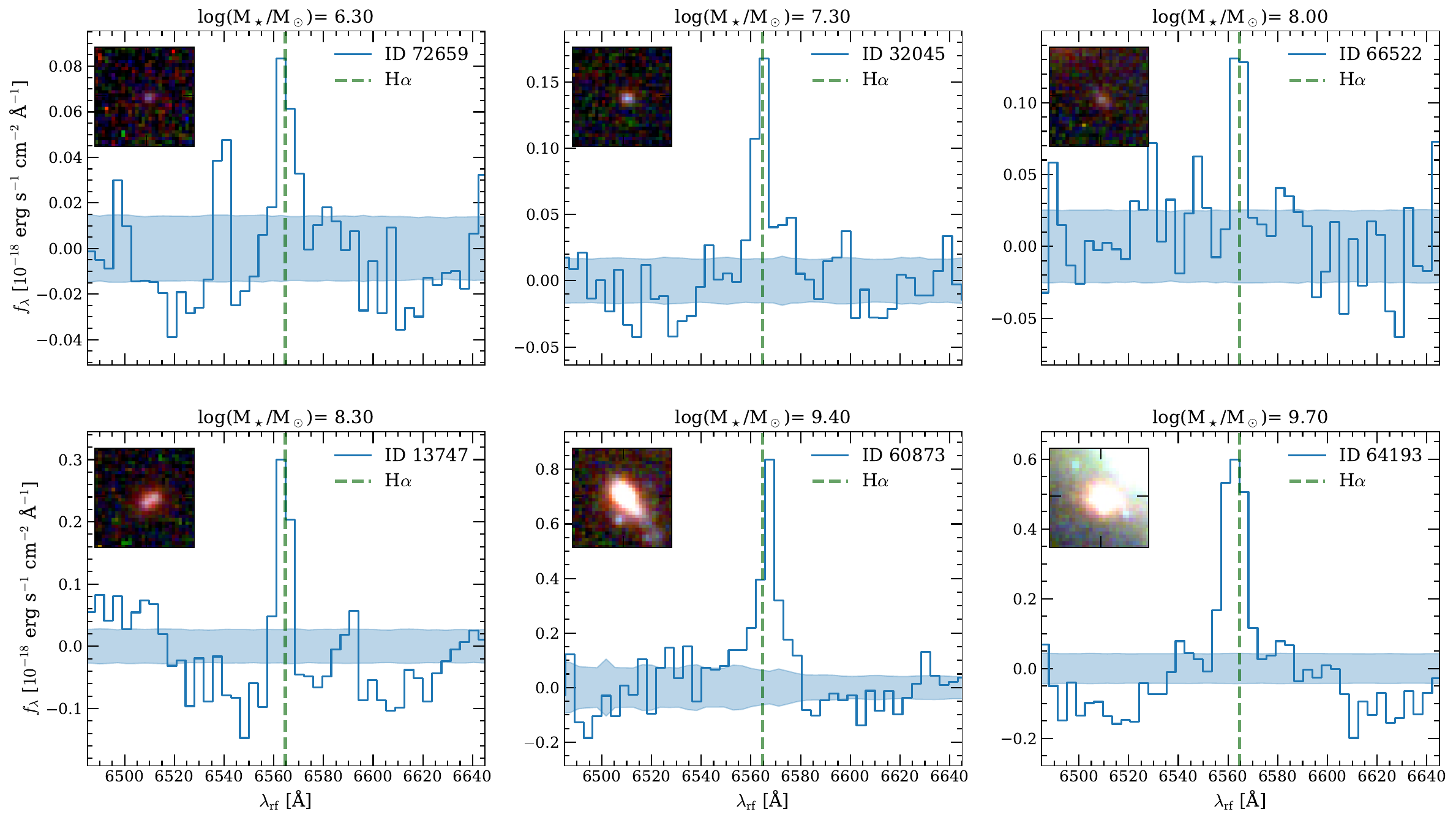}
   \caption{\textbf{F356W grism spectra for six H$\alpha$ emitters in our sample.} On top of each panel we report the stellar mass of the galaxy. The shaded region shows the uncertainty on the flux and the 1.5"$\times$1.5" insets show false-color rest-frame optical RGB images constructed from NIRCam F115W/F200W/F356W. The green and brown lines highlight the H$\alpha$ $\lambda6564.6$ and [\ion{N}{ii}] $\lambda \lambda 6549.9, 6585.4$ wavelengths, respectively.}
   \label{fig:spectra}
\end{figure*}

\subsection{H$\alpha$ emitters in ALT}
\label{sec:Ha_emitters}
The H$\alpha$ line is covered by the ALT observations in the redshift range of $z = 3.7- 5.1$. From the ALT catalog, we find 615 galaxies with H$\alpha$ emission at SNR $\gtrsim 3.5$. Fig. \ref{fig:spectra} shows example spectra together with RGB images of H$\alpha$ emitters at different stellar masses in our catalog.

Since we are interested in star-forming galaxies, we removed 10 sources showing a broad or suspected broad H$\alpha$ component ($v_{\rm FWHM} > 1000$ km/s) from our catalog (N=605), see Table 4 of \citet{Matthee+2024_env} for details on these objects.

\begin{figure}
   \centering
   \includegraphics[scale = 0.45]{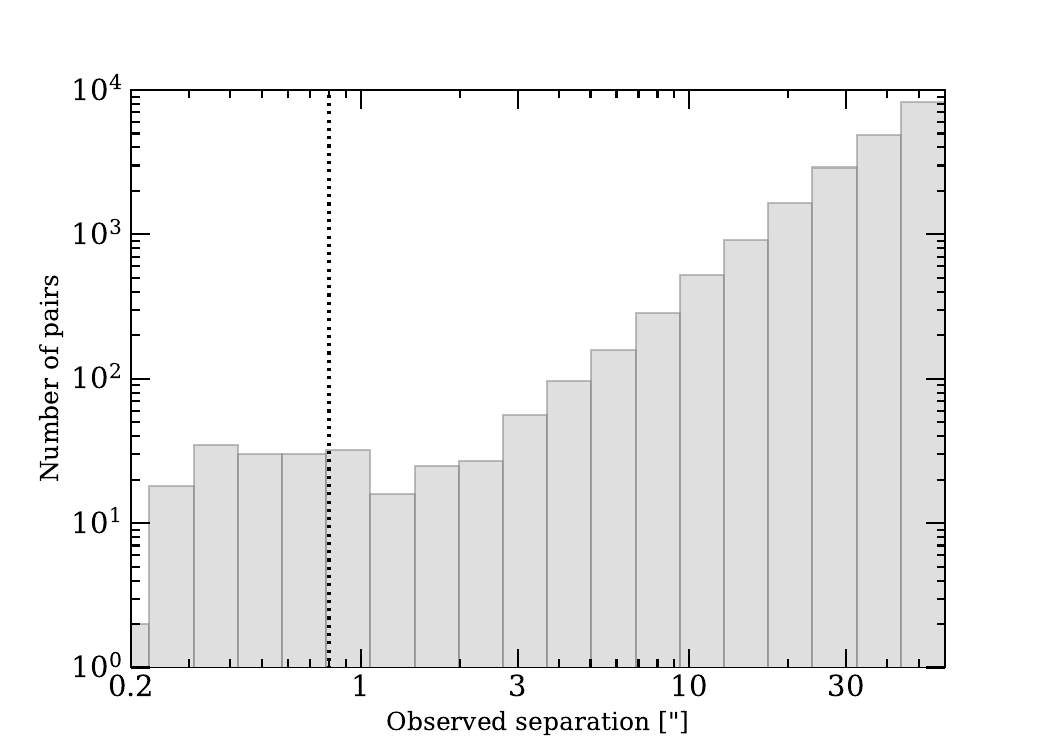}
   \caption{\textbf{Projected separation between all pairs of galaxies} in our sample with $\Delta v < 1000$ km s$^{-1}$. Dashed vertical line at 0.8", which corresponds to $\sim 5.5$ kpc at $z=4.3$, is our choice for the maximum angular separation among components belonging to the same system. }
   \label{fig:separation}
\end{figure}

In some cases, as a consequence of the adopted deblending parameter, galaxies are artificially split into multiple components during the \texttt{SourceExtractor} run. To avoid accounting multiple times for the same system, whose H$\alpha$-flux is blended in the grism data, we adopted a procedure similar to \citet{Matthee+2023_eiger} and grouped sources below a certain angular separation and distance in redshift together. Specifically, we considered sources with $|\Delta v| < 1000$ km/s and angular separation $< 0.8$\arcsec ($\sim 5.5$ kpc at $z = 4.3$) as belonging to the same system and merged their physical properties (e.g. stellar masses and fluxes). This way we ensure that our selection criteria can be reproduced in galaxy simulations and that we are grouping together systems that belong to the same halo, for reference, the virial radius of a $\sim 10^{11} \; \mathrm{M_\odot}$ halo at $z=4.5$ is $\sim30$ kpc. The angular separation was chosen based on that of \citet{Covelo-Paz+2025}, where the authors used an angular separation $< 0.6$ arcsec ($\sim 4.1$ kpc at $z = 4.3$) for their NIRCam grism data, below which issues with blending may impact measurements. Since we are considering objects in a lensed field, we multiply this separation value by the square root of the median magnification factor of the ALT H$\alpha$ sample, $\mu = 1.95$. Figure \ref{fig:separation} shows the pair separation distribution function of the H$\alpha$ emitters in ALT. All the above conditions lead to a sample of 512 objects, of which 61 ($\sim$ 12\%) are multi-component galaxies. 

Finally, given the heterogeneity in the methods applied to identify our sample, as well as the wavelength-dependent sensitivity of our dataset, we decided to apply a conservative cut in the H$\alpha$ flux and magnification to obtain a robust sample. Thus, we only consider galaxies with observed H$\alpha$ fluxes larger than 10$^{-18}$ erg s$^{-1}$ cm$^{-2}$, which allows for a 5$\sigma$ line sensitivity in ALT\footnote{Using Eq. \ref{eq:HatoSFR}, the observed H$\alpha$ flux $\mathrm{f_{H \alpha}} = 10 ^{-18} \mathrm{\; erg \; s^{-1} \; cm^{-2}}$ translates to H$\alpha$ star formation rates of $\mathrm{SFR _{H\alpha}} = 0.7 \; \mathrm{M _\odot \; yr ^{-1}}$ at redshift $z = 5$.}. This condition removes 74 galaxies from our sample. Moreover, to avoid being strongly affected by uncertainties in the magnification model, we restrict our sample to $\mu \leq 2.5$, which excludes 144 galaxies from our sample. We end up with a robust H$\alpha$ sample of 316 objects with SNR $\gtrsim 5$, of which 43 ($\sim 14$ \%) consist of multiple components grouped together, i.e. multi-component galaxies. Although adopting a distance threshold to identify multi-component galaxies may cause close pairs of compact merging systems to be treated as a single source, multi-component galaxies constitute only 14\% of the robust sample, and close mergers are expected to represent only a subset of these. Therefore, we do not expect the presence of close mergers in our sample to impact our results.

\begin{figure}
   \centering
   \includegraphics[scale=0.42]{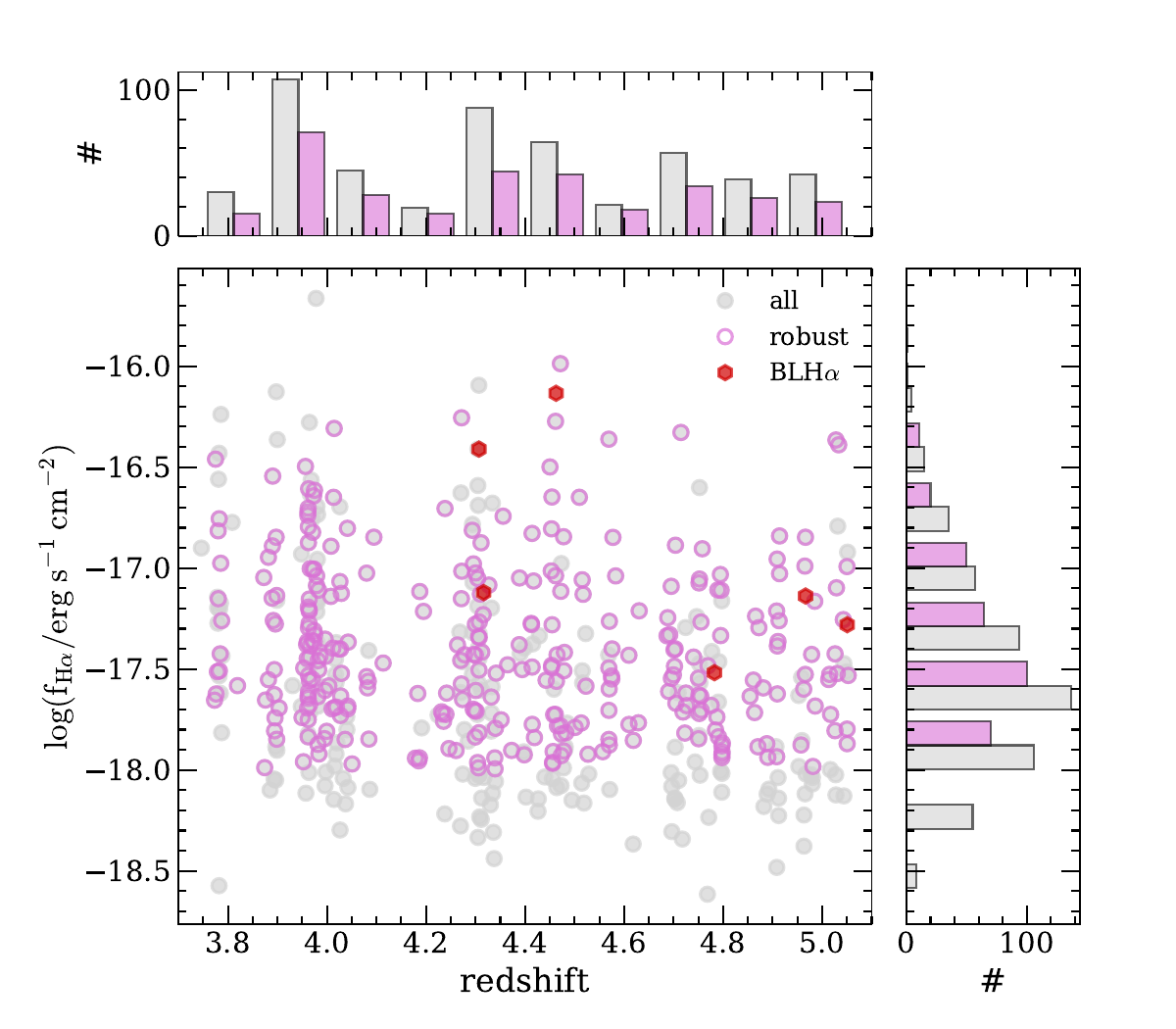}
   \caption{\textbf{Observed H$\alpha$ fluxes as a function of redshift.} In gray is the parent H$\alpha$ sample (N = 512), while in pink we highlighted the robust sample (N = 316, observed H$\alpha$ flux $> 10^{-18}$ erg s$^{-1}$ cm$^{-2}$ and $\mu \leq 2.5$). Red hexagons highlight the 6 confirmed Broad Line H$\alpha$ emitters (BLH$\alpha$), with $v_{\rm FWHM} > 1000$ km/s, presented in \citet{Matthee+2024_env}. The histograms show the redshift and H$\alpha$ flux distributions for both the parent and robust samples.
   }
   \label{fig:sample}
\end{figure}

Figure \ref{fig:sample} shows the observed H$\alpha$ fluxes, not corrected for magnification, as a function of the spectroscopic redshift together with the redshift distribution of the H$\alpha$ emitters. With our robust sample, we probed the entire redshift range of HAEs in ALT.

\section{Methods}
\label{sec:methods}
\subsection{SED fitting}
The physical properties of all galaxies in the ALT survey are estimated using the SED-fitting model \texttt{Prospector} \citep{Leja+2017,Leja+2019, Johnson+2021}. All available photometry (JWST and HST, 27 bands) is used for the SED fitting with the exception of filters including the Ly$\alpha$ emission line and blue wards, due to the impact of the intergalactic medium. A detailed description of the adopted parameters can be found in \citetalias{Naidu+2024_alt}. In short, we used stellar population models \citep{ConroyGunn2010_software, Conroy+2009, Conroy+2010} that assume a \citet{Chabrier2003} IMF and self-consistently include nebular emission modeled with \texttt{Cloudy} \citep{Ferland+2017} grids described in \citet{Byler+2017}. The parameters we fitted for include a non-parametric SFH, the total stellar mass, stellar and gas-phase metallicities, nebular emission parameters and a flexible dust model \citep{Kriek+Conroy2013}.

For the SFH we adopt the “bursty continuity” prior of \citealt{Tacchella+2022}, with time bins logarithmically spaced up to a formation redshift of $z = 20$. The stellar and gas phase metallicities are not tied to each other and span $\log{(\mathrm{Z/Z_{\rm{\odot}}})}=-2$ to 0.19 and -2 to 0.5 respectively. Dust attenuation has been modeled using a two-component dust attenuation model with flexible attenuation curve \citep{CharlotFall2000}. This model accounts for a birth-cloud component that attenuates nebular and stellar emission from stars formed in the last 10 Myr and a diffuse component that has a variable attenuation curve \citep{Kriek+Conroy2013} and attenuates stellar and nebular emission from the galaxy (details on dust attenuation assumptions explained in \citealt{Naidu+2024_alt, Tacchella+2022}).

The dust correction at the H$\alpha$ wavelength was derived by generating posterior SEDs by including the dust parameters and then by setting them to zero. 

The emission line fluxes measured from the grism data are not included in the fitting procedure, as they are covered by medium-band photometry. For almost all sources, the photometry also covers the optical continuum which is free from emission-line contamination in various broad- and/or medium-band filters. 

\subsection{SFR estimates}
In this work, we focus on star formation on short timescales ($\leq 10$ Myr), for which the nebular emission from H$\alpha$ line is a particularly effective tracer.

The conversion from intrinsic H$\alpha$ luminosity to SFR is typically derived using spectral synthesis models and depends on several assumptions, including IMF, star formation history, and stellar metallicity \citep{Tacchella+2022_Halocal}. For example, stellar atmospheric temperatures — and hence the ionizing-photon output — depend on metallicity owing to line blanketing, i.e. the absorption of high-energy radiation by metal lines in stellar atmospheres. A widely used calibration is that of \citet{Kennicutt1998b} (see also \citealt{KennicuttEvans2012}) which assumes a constant SFH over $t=100$~ Myr and solar metallicity, based on the typical conditions in galaxies in the low-redshift Universe. While this calibration is commonly applied to high-redshift galaxies, it should be noted that such systems often have sub-solar metallicity, which may affect its accuracy.

To compute the SFRs of the galaxies in our sample we used dust-corrected H$\alpha$ luminosities. The relation between intrinsic H$\alpha$ luminosity ($\mathrm{L^{int}_{H\alpha}}$) and SFR is as follows:

\begin{equation}
    \rm log\left(\frac{SFR}{M_\odot \: yr^{-1}}\right) = log\left(\frac{L^{int}_{H\alpha}}{erg \: s^{-1}}\right) + C
    \label{eq:HatoSFR}
\end{equation}

\noindent where C is a metallicity and IMF dependent conversion factor, see discussion in \citet{McClymont+2025} and \citet{Kramarenko+2025_sphinx}. Following \citet{Theios+2019} \citep[see also][]{Clarke+2024} we adopt $C = - 41.59$, suitable for high-$z$ galaxies \citep[see e.g.][]{Shapley+2023, Pollock+2025}. Estimates for the SFR are highly dependent on the assumed conversion factor which, in turn, depends on the assumed metallicity. By adopting a standard calibration from \citet{KennicuttEvans2012}, we would have predicted a factor $\sim 2 \times$ higher SFR values than our current estimates. This is because low-metallicity stars are hotter and have reduced line blanketing, both effects leading to greater ionizing photon output and, consequently, more H$\alpha$ emission per unit SFR.

Equation \ref{eq:HatoSFR} assumes that the LyC escape fraction ($f_{\rm esc}^{\rm LyC}$) is equal to zero, i.e. all ionizing photons are absorbed by hydrogen in the ISM of the galaxy. Although this may not be true for all galaxies in our sample, only very large $f_{\rm esc}^{\rm LyC}$ would affect our results. \citet{Mascia+2023_fesc} and \citet{Giovinazzo+2025} analyzed the JWST data ($4.5 \leq z \leq 8$) on A2744 finding $f_{\rm esc}^{\rm LyC} \sim 0.1$ or lower, corresponding to a correction factor of only $\sim 1.1$.

\begin{figure}
   \centering
   \includegraphics[scale = 0.5]{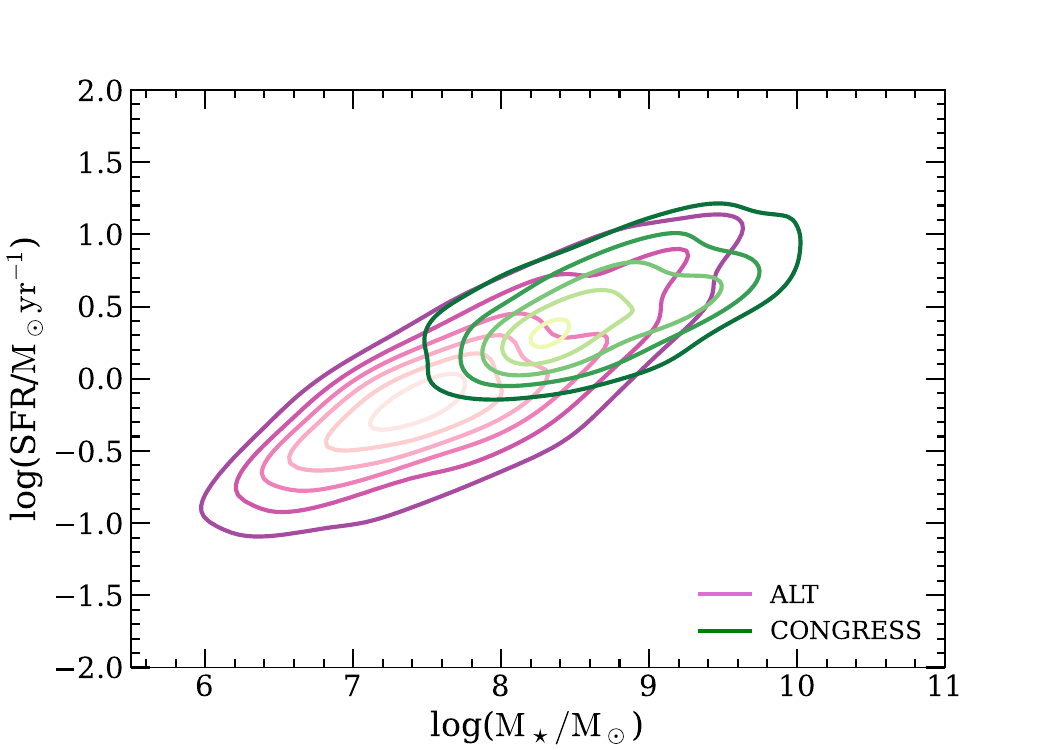}
   \caption{\textbf{The probed parameters space in the SFR-M$_\star$ plane.} The parameter space probed by ALT (pink) compared to that of the CONGRESS survey (green). Both surveys cover H$\alpha$ emission in the redshift range $z=3.7-5.1$. CONGRESS \citep{Egami+2023jwst.prop,Covelo-Paz+2025} employs F356W grism in the GOODS-North field and captures galaxies with a median stellar masses higher than those from ALT, while ALT extends the parameter space approximately 2 orders of magnitude lower in stellar mass.}
   \label{fig:contours}
\end{figure}

Fig. \ref{fig:contours} shows the SFR-M$_\star$ plane for $z=3.7-5.1$ probed by ALT compared to that of the CONGRESS \citep{Egami+2023jwst.prop, Covelo-Paz+2025} survey. CONGRESS employs F356W grism in the GOODS-North field to perform a blind search for H$\alpha$ emitters in the redshift range $z=3.7-5.1$. Since ALT has significantly longer exposure time and observes galaxies behind the A2744 lensing cluster, we are able to access lower mass galaxies that extend the parameter space down a factor $\approx30$ to $\rm M_\star \sim 10^6 \; M_\odot$.  

\subsection{Bayesian model}
\label{sec:bayesian_model}
We implemented a Bayesian model, using the Markov Chain Monte Carlo (MCMC) method\footnote{Specifically, we use the \texttt{autoemcee} Python package (\url{https://johannesbuchner.github.io/autoemcee/readme.html}).} to fit the relationship between SFR and M$_\star$, while properly accounting for controlled observational selection effects. In this model, we parametrize the SFMS as a linear relation between SFR and M$_\star$ in log-log space. We include the intrinsic Gaussian scatter of the relation, which may be mass-dependent. Although previous works found evidence for a redshift-dependent flattening in the SFMS at high masses, leading to a different parameterization of this relation \citep[e.g.,][]{Lee+2015,Popesso+2023}, we adopt a linear relation since all galaxies in our sample lie below the characteristic mass $\sim 10^{11} \; \rm M_\star$ of the turnover at $z \sim 4.5$. The fitted relation in the log-log space is: 
\begin{equation}
\label{eq:sfms_linear}
    {\rm log\left( \frac{SFR}{M_\odot \; yr^{-1}}\right)} = \alpha \times {\rm log\left(\frac{M_\star}{10^{10.5} \; M_\odot}\right)} + \beta + \mathcal{N}(0 \, , \, \sigma_{int} ^2) .
\end{equation}
\noindent In the following, $\alpha$ = ms\_slope and $\beta$ = ms\_norm, which is the intercept at $\rm  log(M_\star/M_\odot) = 10.5$. We use the notation $\mathcal{N}(0 \, , \, \sigma_{int} ^2)$ to indicate a Gaussian distribution with zero mean and variance $\sigma_{int}^2$. The standard deviation, i.e. intrinsic Gaussian scatter, $\sigma_{int}$, is assumed to be related to the stellar mass following a linear relation:
\begin{equation}
\label{eq:int_scatter}
\sigma_{int}  = a \times \; {\rm log\left(\frac{M_\star}{10^{10.5} \; M_\odot}\right)} + b \;.
\end{equation}
\noindent The total observed scatter of the relation is defined as $\sigma = \sqrt{\sigma_{obs}^2 + \sigma_{int} ^2}$ where $\sigma_{obs}$ accounts for the observational uncertainties. There are four key parameters in this model: (ms\_slope, ms\_norm) and \textit{(a,b)}. In Sect. \ref{sec:non-gaussian} we test the effect of a non-Gaussian scatter around the SFMS, which accounts for the presence of a low-SFR tail at fixed stellar mass.

\begin{figure}[h]
    \centering
    \begin{subfigure}{0.4\textwidth}
        \centering
        \includegraphics[width=\textwidth]{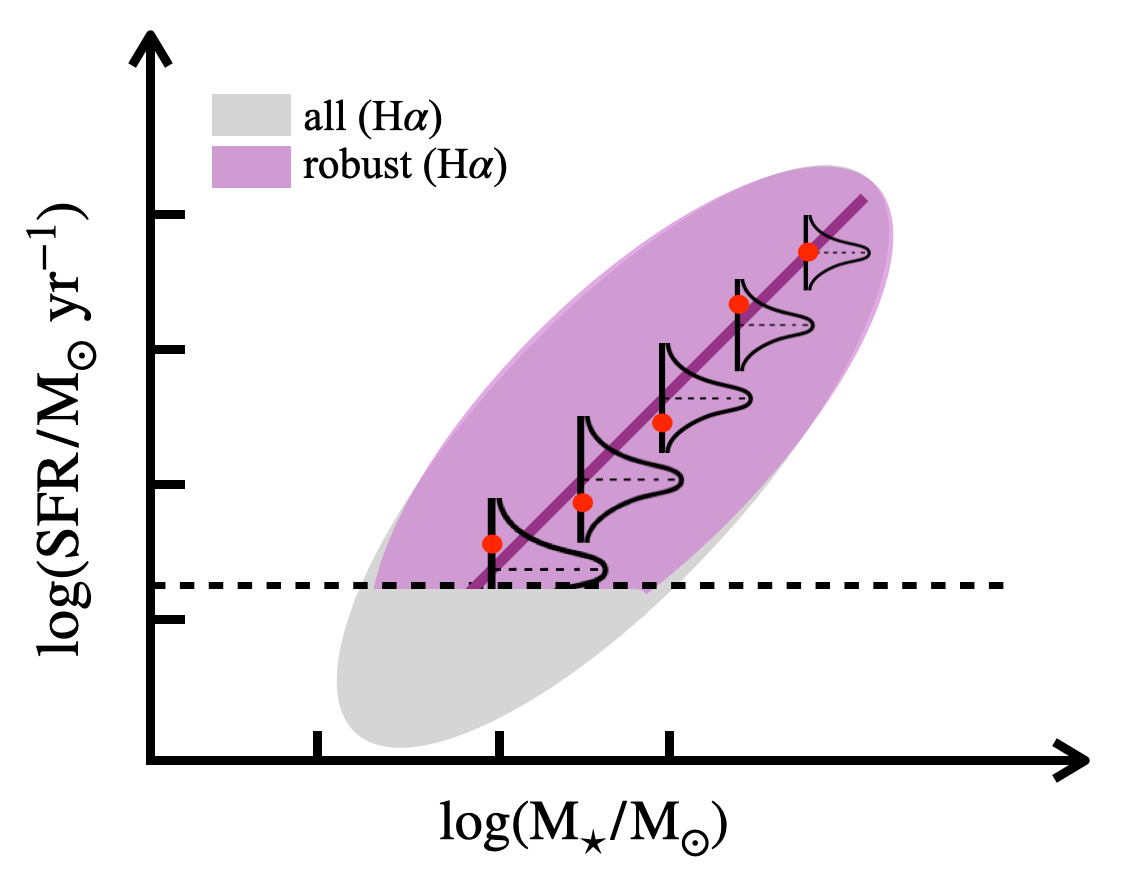}
        \label{fig:sub1}
    \end{subfigure}
    \hfill
    \begin{subfigure}{0.4\textwidth}
        \centering
        \includegraphics[width=\textwidth]{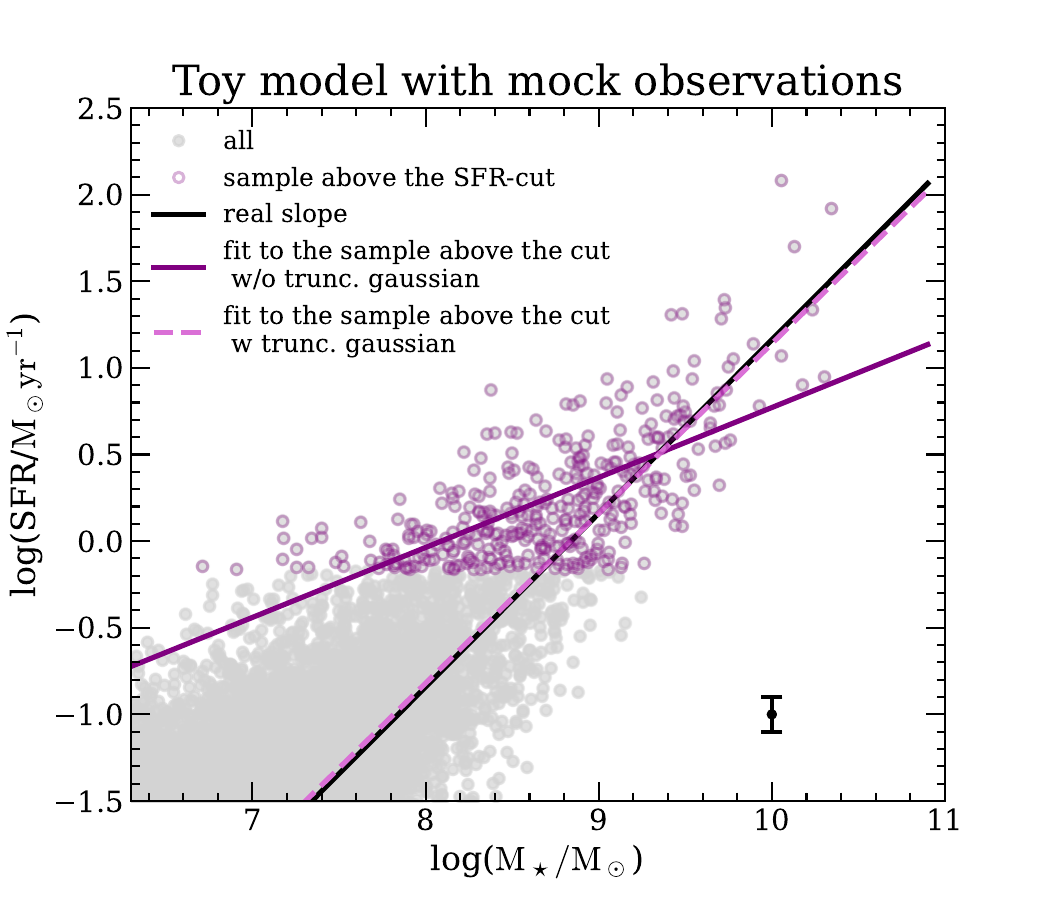}
        \label{fig:sub2}
    \end{subfigure}
    \caption{\textbf{Top: Illustration of the model used to describe the observed SFMS.} The gray cloud represents all the HAEs in ALT, while the pink one is the observed sample after applying the selection criteria for robustness. Red points are the observed SFR, whose probability at fixed stellar mass is modeled using a truncated normal distribution that can be mass dependent. The normal distributions are centered in the SFR values predicted using the observed stellar masses and Eq. \ref{eq:sfms_linear} whose parameters are explored within the MCMC sampling. \textbf{Bottom: Toy model results.} Mock observations generated sampling the stellar mass function from \citet{Weibel+2024} with n\_samples = 30000. Gray dots show the entire sample, purple circles are galaxies with SFR $> 0.7 \; \mathrm{M_\odot/yr}$. The black solid line is the fit to the entire sample, while pink dashed and purple solid lines respectively show the fits with and without truncated normal distributions in the Bayesian model. The error bar in the bottom right is the 0.1 dex uncertainty on the SFR for each mock observation, corresponding to the average star formation rate uncertainty derived from observational data.}
    \label{fig:cartoon-mock}
\end{figure}

We compute the posterior likelihood using the probability function of the star formation rate ($p$). In particular, we model SFR distributions using truncated normal distribution functions centered on the SFR values calculated from Eq. \ref{eq:sfms_linear}, using the observed M$_\star$, with the total sigma ($\sigma$) as the standard deviation. The uncertainties on SFRs are assumed to be Gaussian and enters as $\sigma _{obs}$ in the total sigma. Each Gaussian distribution is truncated to mimic the SFR-limited selection of our sample, resulting in the selection-corrected probability $p$. The truncation is defined using Eq. \ref{eq:HatoSFR} with f$\rm _{H\alpha} = 10^{-18} \; erg \; s^{-1} \; cm^{-2}$ at fixed $z=5$, which results in a conservative SFR cut in our sample. We refer to this model, schematically illustrated in Fig. \ref{fig:cartoon-mock} (top panel), as our reference model. 

To validate our model, we generate mock observations sampling the stellar mass function from \citet{Weibel+2024} (n\_samples = 30000) and the SFR-M$_\star$ relation with known input parameters (ms\_slope = 1, ms\_norm = 1.66 at $\mathrm{log(M_\star/M_\odot)}= 10.5$, a = - 0.05 and b = 0.3 dex). From the total sample, we define the robust sample as the one with SFR $> 0.69 \; \mathrm{M_\odot/yr}$ -- derived from Eq. \ref{eq:HatoSFR} with f$\rm _{H\alpha} = 10^{-18} \; erg \; s^{-1} \; cm^{-2}$ and $z=5$. Based on the mean value of the uncertainties on the observed robust sample, we assumed Gaussian uncertainties for SFRs of 0.1 dex. Then, we fit the robust mock sample with and without truncated Gaussian distributions in our Bayesian model. Figure \ref{fig:cartoon-mock} (bottom panel) shows that without taking into account the selection effect the cutoff has on the sample, we would estimate a shallower slope of ms\_slope $=0.40\pm 0.02$ and, as a consequence, a smaller normalization (purple solid line). The slope, normalization and scatter-related parameters are recovered (ms\_slope = $0.98^{+0.10}_{-0.08}$, ms\_norm = $1.63^{+0.11}_{-0.11}$, a = $-0.04 ^{+0.02}_{-0.02}$, b = $0.30^{+0.04}_{-0.04}$) once we account for the SFR cutoff and use truncated Gaussian distributions (pink dashed line). 

We used this log-likelihood function as the objective function for MCMC parameter exploration, identifying the combinations of parameters (ms\_slope, ms\_norm, $\; a, \; b$) that best describe the observed sample by maximizing likelihood.

\section{Results}
\label{sec:results}

\subsection{Dust attenuation}
\label{sec:dust_att}

\begin{figure}
   \centering
   \includegraphics[scale=0.5]{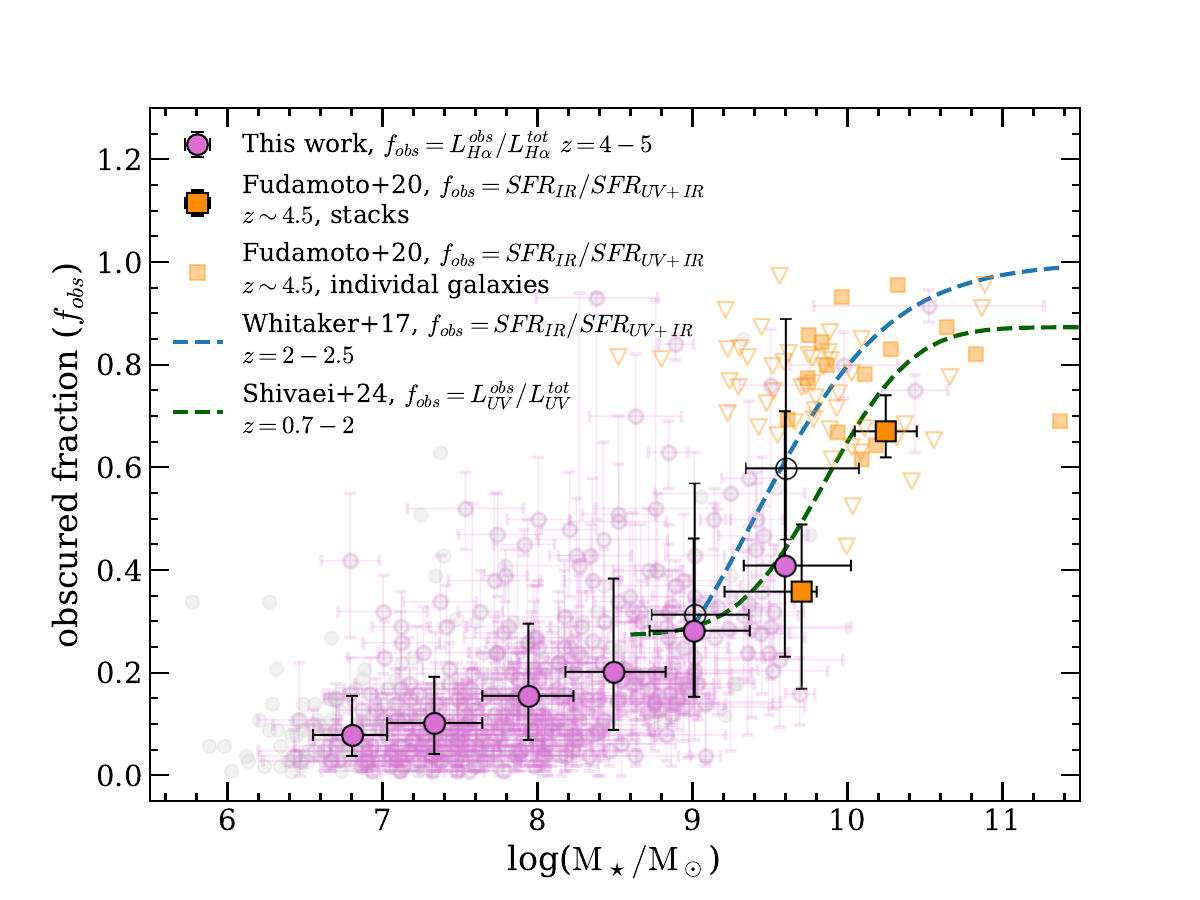}
   \caption{\textbf{Obscured star formation as a function of stellar mass.} The fraction of dust attenuated H$\alpha$ emission, i.e. the ratio between the obscured and total H$\alpha$ luminosity, as a function of stellar mass for the ALT sample. In gray is the parent H$\alpha$ sample, while highlighted in pink is the robust one. Median values in each mass bin are shown as filled circles. Empty markers show obscured H$\alpha$ fraction once we account for enhanced dust attenuation in high-mass galaxies (see Sect. \ref{sec:test_dustatt}). 
   As a comparison, we show the fraction of obscured star formation for galaxies at $2<z<2.5$ from \citet{Whitaker+2017} (blue), at $0.7 < z < 2$ from \citet{Shivaei+2024} (green) and at $z\sim 4.5$ ALPINE galaxies from \citet{Fudamoto+2020_alpine} (orange), with squares showing individual FIR continuum detections at $4<z<5$ and triangles the 3$\sigma$ upper limits for IR non detections. Filled squares show stacks in 2 mass bins at $z \sim 4.5$.}
   \label{fig:dust_corr}
\end{figure}

Figure \ref{fig:dust_corr} shows the H$\alpha$ obscured fraction, estimated using \texttt{Prospector}, as a function of stellar mass for the ALT robust sample. Despite significant scatter in the relationship, we observe the trend of increasing dust attenuation with stellar mass. The H$\alpha$ is heavily obscured ($f_{obs}$ > 50\%) above the stellar mass of $\sim 6 \times 10^9 \; \mathrm{M_\odot}$, however most (95\%) of the galaxies in our robust sample have an obscured fraction of H$\alpha < 50 \% $. Interestingly, due to the large scatter in $f_{obs}$ at all masses, there are galaxies with high dust obscuration even down at $\sim 4 \times 10^8 \; \mathrm{M_\odot}$. 

In Fig. \ref{fig:dust_corr} we show the fraction of obscured star formation (i.e. the fraction of star formation activity observable from IR continuum emission relative to its total amount) at $2<z<2.5$ estimated by \citet{Whitaker+2017} using {\it Spitzer}/MIPS observations of a mass complete sample at $\mathrm{log(\mathrm{M_\star/M_\odot})} \gtrsim 9$. The authors find that $f_{obs}$ is highly mass dependent, with more than 80\% of star formation being obscured at $\mathrm{log(\mathrm{M_\star/M_\odot})} \gtrsim 10$ and that this relation does not evolve with redshift. In \citet{Whitaker+2017}, the authors discussed that for $\mathrm{M_\star} < 10^{10} M_\odot$ the estimated IR luminosity systematically changes depending on the FIR SED template adopted. Their default template does not include an evolution of the dust temperature with redshift. More recently, \citet{Shivaei+2024} studied the fraction of obscured UV emission using data from the MIRI imaging survey SMILES \citep{Alberts+2024_smiles, Rieke+2024_smiles} at $z=0.7-2$. \citet{Shivaei+2024} confirm a redshift-independent strong correlation between the dust-obscured fraction and stellar mass. Although the two relationships are in relatively good agreement, \citet{Shivaei+2024} argue that the single FIR SED assumption in \citet{Whitaker+2017} might have led to an overestimation of their IR luminosity values at lower masses and, as a consequence, of their obscured fraction. Furthermore, the different covered areas and the sensitivity of SMILES and {\it Spitzer}/MIPS may also play a role in the observed discrepancies between these two studies. SMILES is more sensitive to less obscured galaxies at fixed mass and covers a smaller area compared to {\it Spitzer}/MIPS. 

Finally, we show the fraction of obscured star formation for ALPINE galaxies \citep{Fudamoto+2020_alpine}. In particular, we show $f_{obs}$ for individual galaxies with and without IR detection and the results of stacks in two mass bins ($\mathrm{M_\star > 10^{10} \; M_\odot}$ and $\mathrm{M_\star < 10^{10} \; M_\odot}$). The stacks show a lower obscured fraction compared to \citet{Whitaker+2017}, but relatively good agreement with \citet{Shivaei+2024}. However, individual sources with IR detection show good agreement with both relations, highlighting an obscured star formation higher than 60\%.

Notably, we find that the fraction of the obscured star formation rate of our robust sample is in agreement with the trend of \citet{Shivaei+2024} for $\mathrm{log(M_\star/M_\odot)} = 9-10$.

\subsection{The SFR-M$_\star$ relation} 
\begin{figure}
   \centering
   \includegraphics[scale=0.5]{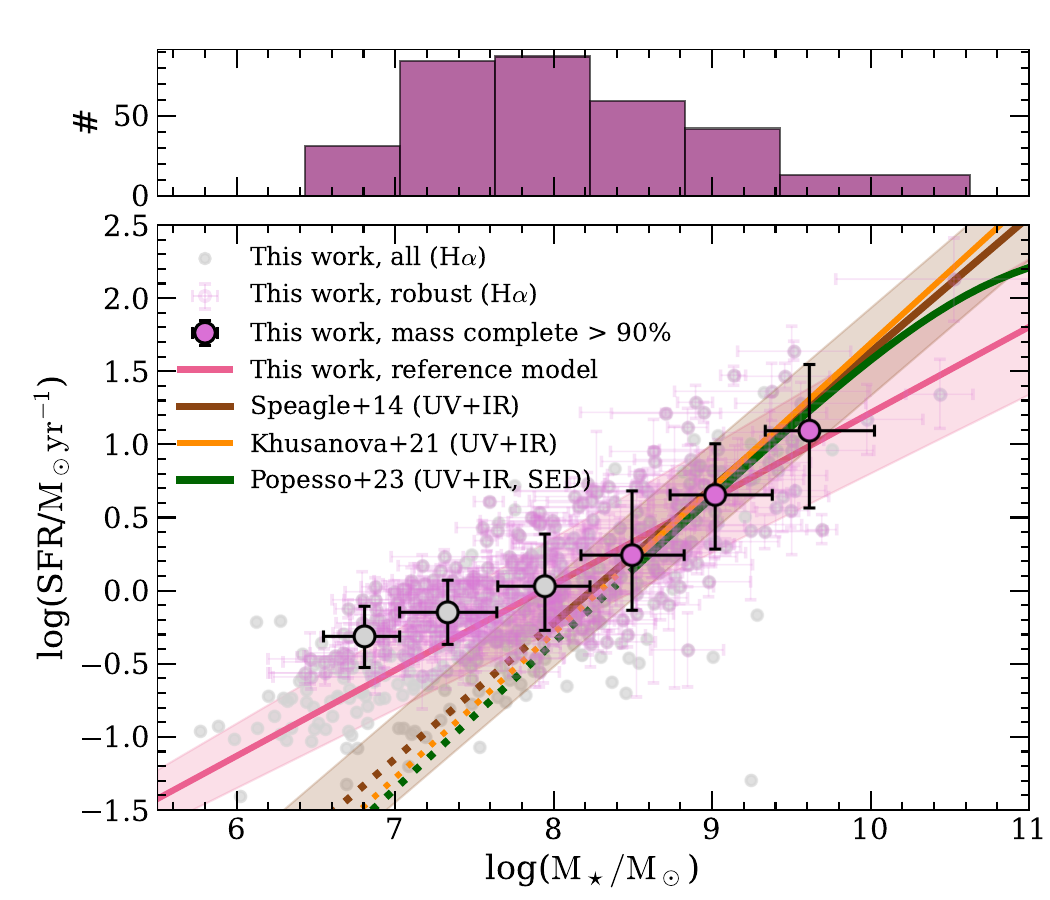}
   \caption{\textbf{The SFR-M$_\star$ relation at $\mathbf{3.7 < z < 5.1}$.} In gray is the entire ALT H$\alpha$ sample, while highlighted in pink is the robust subsample. Circles show the median value when binning in stellar mass (upper panel). We are $>90\%$ mass complete for galaxies with $\rm log(M_\star/M_\odot) > 8.5$ (filled circles). For reference, we show SFMS fit by \citet{Speagle+2014} (together with its 0.3 dex scatter) and \citet{Popesso+2023}, both from a compilation of literature observations, and \citet{Khusanova+2021}, from ALMA-ALPINE observations. Dotted lines mark extrapolations to these relations based on their lower limits on M$_\star$. Solid pink line and shaded region denote respectively the best-fit line and 1$\sigma$ interval about the SFMS using the reference model that accounts for the selection function (see Sect. \ref{sec:bayesian_model}).}
   \label{fig:sfr-mstarALT}
\end{figure}

Figure \ref{fig:sfr-mstarALT} shows the SFMS at $4<z<5$ with SFRs derived from H$\alpha$ and that spans a mass range $\rm log(M_\star/M_\odot) = 6.43 - 10.57$. We show both the entire (gray) and robust (pink) ALT H$\alpha$ samples, together with the median values when binning in stellar mass. The errors on the medians come from bootstrapping with 300 realizations with replacement (16th and 84th percentiles) and bin sizes are chosen to ensure at least
 13 galaxies in each bin. For the robust sample, we reach mass completeness above $90\%$ for galaxies with $\rm log(M_\star/M_\odot) > 8.5$ (see Appendix \ref{app:completeness}), shown with filled circles for the median values. For reference, we also show the SFMS trend from \citet{Speagle+2014} (together with its measured 0.3 dex scatter), \citet{Popesso+2023} and \citet{Khusanova+2021}. The trend from \citet{Speagle+2014}, constrained from $z \sim 0$ to $z\sim5$, is the result of a compilation of literature studies. Their galaxy sample has stellar masses down to $\rm 10^8 \; M_\odot $ and SFRs derived from the rest-frame UV continuum. \citet{Popesso+2023}, compiled more recent literature data to investigate the redshift evolution ($0 < z < 6$) of the SFMS. Most SFR estimates are based on UV+IR data or are derived through SED fitting techniques based on UV+IR from ground-based telescopes, HST and \textit{Spitzer}. They probed the stellar mass range from $\rm 10^{8.5} \; M_\odot $ to $\rm 10^{11.5} \; M_\odot $, highlighting the presence of a bending in the relation for large stellar masses, $\rm log(M_\star/M_\odot) > 10.5$. Finally, we show the SFMS fit from \citet{Khusanova+2021}, based on observations from ALMA-ALPINE \citep{Faisst+2020_alpine, Bethermin+2020_alpine, LeFevre+2020_alpine}. These galaxies, at $4<z<5$, have masses in the range $\rm 8.5 < log(M_\star/M_\odot) < 11$ and SFR estimates derived from rest-frame UV+IR information. All the above trends from literature studies, that based their SFR estimates on UV + IR data, show a slope of the SFR-M$_\star$ relation close to unity. 

\begin{table}[h]
\centering
\renewcommand{\arraystretch}{1.3}
\caption{{\bf The median H$\alpha$-based SFRs in bins of stellar mass in the ALT sample at $z=3.7-5.1$.} We list the median values in each bin and errors correspond to the 16,84th percentiles, respectively. We also list the median {\it mass} completeness of our SFR-selected sample in each mass bin, estimated as described in Appendix $\ref{app:completeness}$.}
\begin{tabular}{c|c|c}  
\hline
\makecell{$\rm log(M_\star/M_\odot)$} & \makecell{$\rm log(SFR/M_\odot~ yr^{-1})$} & Mass completeness \\
\hline
$6.8 ^{+0.2}_{-0.3}$ & $-0.31^{+0.20}_{-0.21}$  & 5 \% \\
$7.3 ^{+0.3}_{-0.3}$ & $-0.13^{+0.21}_{-0.22}$  &  28 \%\\
$7.9 ^{+0.3}_{-0.3}$ & $0.04^{+0.36}_{-0.29}$ &  70 \%\\
$8.5 ^{+0.3}_{-0.3}$ & $0.26^{+0.43}_{-0.38}$  &  90 \%\\
$9.0 ^{+0.4}_{-0.3}$ & $0.66^{+0.35}_{-0.36}$  &  98 \%\\
$9.6 ^{+0.4}_{-0.3}$ & $1.09^{+0.45}_{-0.52}$ &  100 \%\\
\hline
\end{tabular} \label{tab:measurements}
\end{table}

\begin{figure*}
   \centering
   \includegraphics[scale=0.5]{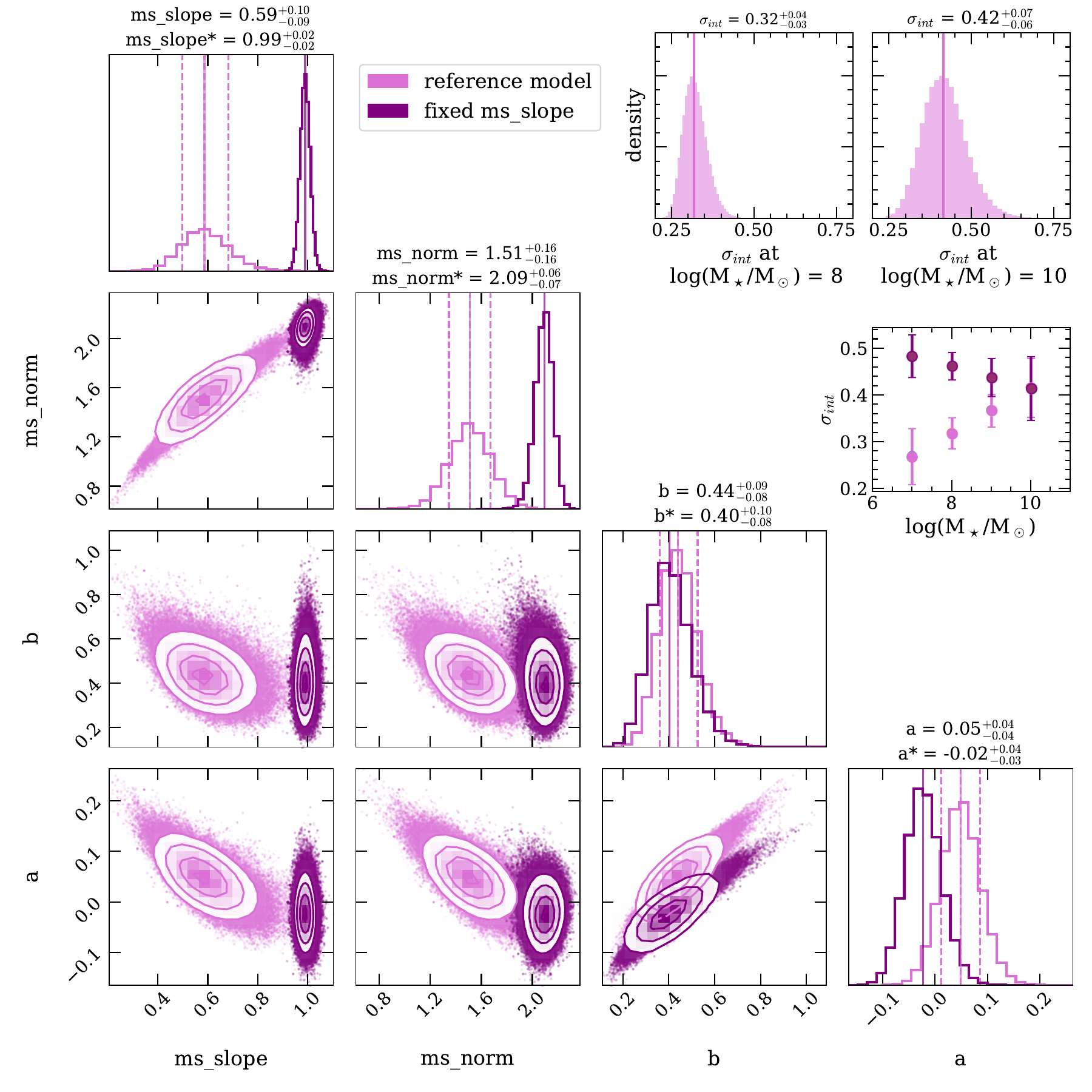}
   \caption{{\bf Output of our reference model.} Corner plot and posterior distributions (pink) of the four key parameters in our reference model: SFMS slope and normalization at $\rm log(M_\star/M_\odot) = 10.5$ (ms\_slope, ms\_norm) and the slope and normalization of the intrinsic scatter \textit{(a,b)}. The contours show the 1$\sigma$, 1.5$\sigma$, and 2$\sigma$ confidence levels. We adopted truncated normal distributions to model the SFRs distribution in our Bayesian model (see Sect. \ref{sec:bayesian_model}). In purple we show the posterior distributions after running the Bayesian model on the robust sample imposing a Gaussian prior (mean=1 and $\sigma = 0.02$) on ms\_slope, referred to as “fixed ms\_slope” in the legend. Above each panel we report medians with uncertainties (16th and 84th percentiles), those with asterisks refer to fixed ms\_slope. The inset panels on the right show the intrinsic scatter distributions at two stellar masses, computed combining the posterior distributions for \textit{a} and \textit{b} with Eq. \ref{eq:sfms_linear}, and the $\sigma_{int}$ evolution with stellar mass for both the reference model and that with fixed ms\_slope.}
   \label{fig:corner_truncnorm}
\end{figure*}

Figure \ref{fig:sfr-mstarALT} shows that for stellar masses above $\rm log(M_\star/M_\odot) = 8.5$ our H$\alpha$ derived main sequence agrees well with that derived from the rest-frame UV and UV+IR continuum at $z=4.5$. Going toward lower stellar masses, we find a flattening in the data points mainly driven by the selection effect on the H$\alpha$ flux, which biases our observed sample (i.e. flux-limited survey). Indeed, because of the widely varying luminosities and faintness of low-mass galaxies, it is difficult to obtain complete samples at a given mass. As a consequence, the selection of galaxies near the flux limit results in a biased sample toward galaxies in the burst phase (Malmquist bias; see \citealt{Reddy+2012}). Moreover, we measure an intrinsic scatter $\sigma _{int} \sim 0.32$ dex at $\rm log(M_\star/M_\odot) = 8$, consistent with the scatter reported by \citet{Smit+2016} at $z \sim 4-5$ and that of \citet{Shivaei+2015} at $z\sim 2$ for higher stellar masses, both based on H$\alpha$ observations. This scatter is slightly larger than the 0.3 dex estimated by \citet{Speagle+2014}, using UV + IR data. Such a dispersion in the robust data sample is not unexpected and can be attributed to bursty star formation histories traced using H$\alpha$ emission line \citep[e.g.,][]{Dominguez+2015_burstiness,Faisst+2019,CaplarTacchella2019, McClymont+2025}. However, as argued by \citet{Shivaei+2015}, without direct measurements of galaxy-to-galaxy variations in attenuation curves and the IMF, differences in the scatter between the SFR(H$\alpha$)–$\mathrm{M_\star}$ and SFR(UV)–$\mathrm{M_\star}$ relations cannot be reliably used to constrain the stochasticity of star formation in high-redshift galaxies.

Using a simple linear fit with \texttt{scipy.curvefit} applied to our robust sample of HAE, we derive a SFMS slope of ms\_slope $ = 0.45 \pm 0.02$, consistent with other JWST/NIRSpec measurements \citep[e.g.,][]{Chemerynska+2024_mzr}, and a normalization of $1.26 \; \pm 0.06 \; \mathrm{M_\odot \; yr^{-1}}$ at $\rm log(M_\star/M_\odot) = 10.5$. However, as the ALT survey is a flux-limited survey, our low-mass sample is biased toward galaxies with higher specific SFRs as illustrated in Fig. $\ref{fig:cartoon-mock}$.

In order to fit the observational sample taking into account its biased nature, we employ the Bayesian model described in Sect. \ref{sec:bayesian_model} assuming uniform prior probability distributions, with each parameter restricted to the range: ms\_slope$=[0.2,1.8]$, ms\_norm$=[0.01,2.4]$, $a=[-0.1,0.3]$, $b=[0.1,1.3]$. We used Bayesian inference through MCMC sampling, using the \texttt{autoemcee} Python package, to explore the parameter space and determine posterior distributions for the parameters. The solid line and shaded region in Fig. \ref{fig:sfr-mstarALT} show the best-fit model and 1$\sigma$ interval when applying our reference model to the robust H$\alpha$ sample from ALT. We list the median SFRs in stellar mass bins in Table $\ref{tab:measurements}$, as well as the estimated stellar mass completeness of the galaxy sample in each mass bin (see Appendix \ref{app:completeness}).

Figure \ref{fig:corner_truncnorm} presents the corner plot for the inferred posterior distributions for the parameters explored in our model: the slope and normalization of the SFMS (ms\_slope, ms\_norm) and the slope and normalization of the intrinsic scatter (\textit{a,b}) -- see Eqs. \ref{eq:sfms_linear} and \ref{eq:int_scatter}. The three panels in the upper right corner show the mass evolution of $\sigma_{int}$, derived using the posterior distributions for \textit{a} and \textit{b} together with Eq. \ref{eq:sfms_linear}. From the corner plot, we observe a strong positive correlation between ms\_slope and ms\_norm, reflecting the slope-intercept degeneracy that arises when fitting linear relationships. Also, a positive correlation between the scatter parameters ($b$ -- $a$), highlighting that intrinsic scatter normalization and its mass dependence are degenerate. Table \ref{tab:table} reports the median value of the MCMC sample distribution and the 16th and 84th percentiles (i.e. 1$\sigma$ confidence interval). The median posterior parameters, which are consistent with the best-fit values, suggest a steeper slope compared to that of \texttt{curvefit} - ms\_slope $=0.59 ^{+0.10}_{-0.09}$ - normalization of $1.44 \pm 0.15 \mathrm{M_\odot \; yr^{-1}}$ at $\rm log(M_\star/M_\odot) = 10.5$. Although the slope is consistent with that reported by \citet{Shivaei+2015} for H$\alpha$ emitters at $z\sim 2$, it is shallower than that derived from UV+IR data \citep[e.g.][]{Popesso+2023}. In the literature, some studies report constant scatter \citep[e.g.,][]{Clarke+2024}, while others find evidence for mass-dependent variation \citep[e.g.,][]{Dominguez+2015_burstiness,Santini+2017, Cole+2025_ceers, Clarke+2025}. In our analysis, we find an intrinsic scatter that slightly increases with stellar mass (inset panels, Fig. \ref{fig:corner_truncnorm}). This is suggested by the positive value assumed by \textit{a}, which controls the mass dependence of the intrinsic scatter. However, the large 1$\sigma$ uncertainty associated with this parameter prevents us from drawing strong conclusions about the mass dependence of the scatter. 

\begin{table*}
    \centering
    \renewcommand{\arraystretch}{1.5}
    \caption{{\bf The SFMS slope, normalization at $\rm log(M_\star/M_\odot) = 10.5$ and intrinsic scatter estimated using the Bayesian model with truncated normal distributions (Sect. \ref{sec:bayesian_model}).} This model is applied to observations (i.e. the ALT robust H$\alpha$ sample), adopting different conversion between H$\alpha$ luminosity and SFR (\citealt{Theios+2019} - fiducial - and \citealt{Kramarenko+2025_sphinx}) and modified dust attenuation corrections and to data from hydrodynamical simulations \citep{Graziani+2020_dustyG, Schaye+2015_eagle, Pillepich+2019_tng50, Rosdahl+2018_sphinx, McClymont+2025}. Also, we report the parameter estimates from the Bayesian model using truncated skewed distributions, applied to the observational robust sample.}
    \begin{tabular}{c|c|c|c|c}
    \hline
    \multicolumn{5}{|c|}{\textbf{ALT H$\alpha$ observations}} \\
    \hline
    & ms\_slope & ms\_norm  & $\sigma_{int}$ at $\rm log(M_\star/M_\odot) = 8$  & $\sigma_{int}$ at $\rm log(M_\star/M_\odot) = 9$\\
    \hline
        {\bf fiducial} -- \citealt{Theios+2019} \textbf{(ref.)} & $0.59^{+0.10}_{-0.09}$ & $1.51 ^{+0.16} _{-0.16}$  & $0.32^{+0.04}_{-0.03}$  & $0.37^{+0.03}_{-0.04}$ \\
        SFR = SFR(L$_{\rm H\alpha}$)\tablefootmark{1}& $0.66 ^{+0.08} _{-0.08}$ & $1.79 ^{+0.14}_{-0.15}$  & $0.35 ^{+0.04}_{-0.03}$ & $0.37 ^{+0.04}_{-0.03}$ \\
        SFR = SFR(L$_{\rm H\alpha}$, EW(H$\alpha$))\tablefootmark{1} & $0.67 ^{+0.06} _{-0.05}$ & $1.95 ^{+0.11}_{-0.11}$  & $0.29 ^{+0.02}_{-0.02}$  & $0.30 ^{+0.02}_{-0.03}$ \\
        modified dust attenuation\tablefootmark{2} & $0.73 ^{+0.11} _{-0.10}$ & $1.79 ^{+0.17}_{-0.18}$  & $0.35 ^{+0.04}_{-0.04}$  & $0.37 ^{+0.04}_{-0.03}$ \\
    \hline
    & ms\_slope & ms\_norm  & $\sigma_{eff}$ at $\rm log(M_\star/M_\odot) = 8$  & $\sigma_{eff}$ at $\rm log(M_\star/M_\odot) = 9$\\
    \hline
        skewed distribution\tablefootmark{3} & $0.63 ^{+0.14} _{-0.13}$ & $1.46 ^{+0.22}_{-0.26}$  & $0.46 ^{+0.11}_{-0.08}$  & $0.49 ^{+0.11}_{-0.07}$ \\
    \hline
    \multicolumn{5}{|c|}{\textbf{Hydrodynamical simulations}} \\
    \hline
    & ms\_slope & ms\_norm  & $\sigma_{int}$ at $\rm log(M_\star/M_\odot) = 8$  & $\sigma_{int}$ at $\rm log(M_\star/M_\odot) = 9$\\
    \hline
        dustyGadget \tablefootmark{a}  & $0.97 ^{+0.02}_{-0.02}$ & $1.86 ^{+0.02}_{-0.02}$ & $0.15 ^{+0.01}_{-0.01}$ & $0.14 ^{+0.01}_{-0.01}$ \\
        EAGLE \tablefootmark{b} & $1.00 ^{+0.02}_{-0.02}$ & $1.90 ^{+0.03}_{-0.03}$ & $0.19 ^{+0.01}_{-0.01}$ & $0.21 ^{+0.01}_{-0.01}$\\
        Illustris-TNG50 \tablefootmark{c} & $1.09 ^{+0.01}_{-0.01}$ & $2.22 ^{+0.02}_{-0.02}$ & $0.13 ^{+0.01}_{-0.01}$  & $0.12 ^{+0.01}_{-0.01}$\\
        SPHINX \tablefootmark{d} & $0.95 ^{+0.07}_{-0.07}$ & $1.71 ^{+0.08}_{-0.08}$ & $0.46 ^{+0.04}_{-0.04}$ & $0.35 ^{+0.03}_{-0.03}$ \\
        THESAN-ZOOM \tablefootmark{e} & $ 0.98^{+0.06}_{-0.06}$ & $ 1.70^{+0.17}_{-0.17}$ & $ 0.48^{+0.03}_{-0.03}$ & $ 0.51^{+0.05}_{-0.05}$ \\
    \end{tabular}
        \tablefoot{\\\tablefoottext{1}{Calibration from \citet{Kramarenko+2025_sphinx} using SPHINX simulations, see Sect. \ref{sec:new_calibrations}.} \tablefoottext{2}{Modified dust attenuation values for high-mass galaxies, see Sect. \ref{sec:non-gaussian}.} \tablefoottext{3}{Fit to the data using truncated skewed distributions in the Bayesian framework, see Sect. \ref{sec:non-gaussian}.} \\ \tablefoottext{a}{Data at $z=4.495$, private communication.}; \tablefoottext{b}{Public data set at $z = 3.98$ and $z = 4.49$ taken from \url{https://eagle.strw.leidenuniv.nl/wordpress/}}; \tablefoottext{c}{public data set at $z =4.66$ taken from \url{https://www.tng-project.org/data/}}; \tablefoottext{d}{Public data set at $z =4.64$ taken from \url{https://github.com/HarleyKatz/SPHINX-20-data}}; \tablefoottext{e}{Data at $z =4.46$, private communication.}}
\label{tab:table}
\end{table*}

\subsection{Comparison with other observations}
\label{sec:comparison_obs}
\begin{figure}
   \centering
   \includegraphics[scale=0.5]{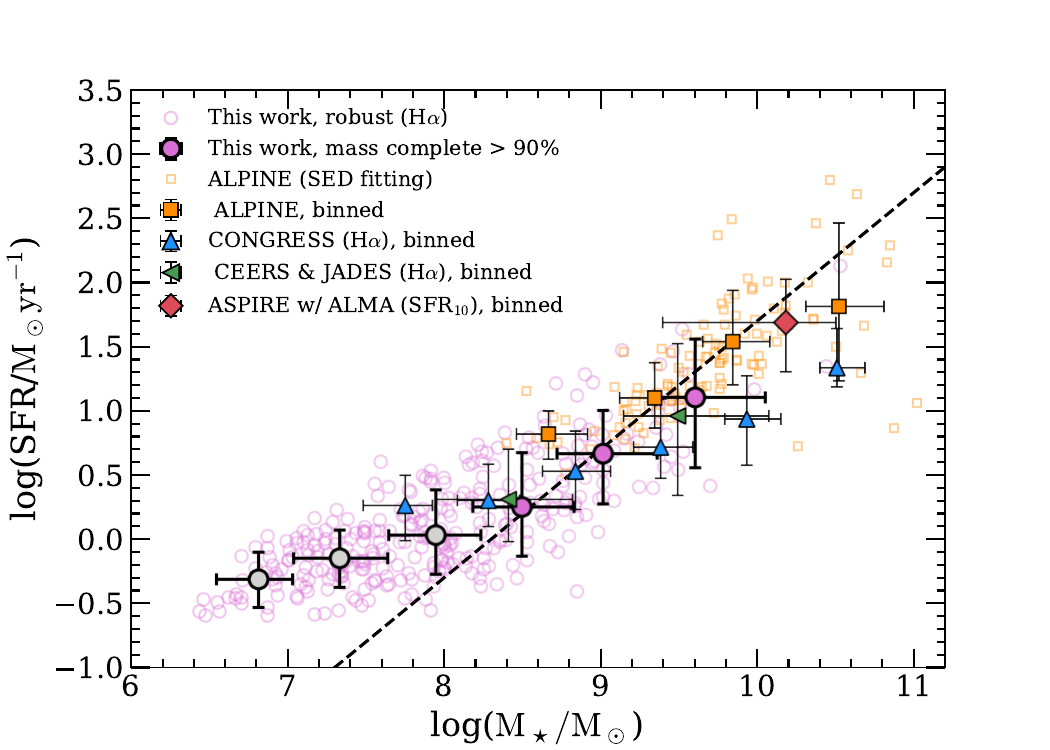}
   \caption{{\bf Comparison with literature studies in the redshift range $4 < z < 5$.} The robust H$\alpha$ sample used in this work and its binned values are in pink circles, blue and green triangles are mass binned observations from CONGRESS \citep[][]{Covelo-Paz+2025} and, CEERS and JADES JWST-surveys \citep[][]{Clarke+2024}, respectively. Orange squares show observations from ALMA-ALPINE \citep[][]{Faisst+2020_alpine}, while the red square is the binned value for the ASPIRE survey which combines JWST and ALMA observations \citep[][]{Sun+2025_aspire}. To guide the eye, we show a linear relation with slope one (dashed black).}
   \label{fig:sfms_literature}
\end{figure}

In Fig. \ref{fig:sfms_literature}, we compare the SFMS of our robust H$\alpha$ sample with that of samples in the literature \citep{Covelo-Paz+2025, Clarke+2024, Faisst+2019, Sun+2025_aspire}. We binned in mass the available observations to compute the medians and 1$\sigma$ uncertainties by bootstrapping, where we resample our galaxies with replacements 300 times. The SFR values in CONGRESS \citep{Covelo-Paz+2025} and JADES+CEERS \citep{Clarke+2024} data have been derived using the same conversion between L$_{\rm H\alpha}$ and SFR as in this work, while those of ALPINE \citep{Faisst+2020_alpine} and ASPIRE \citep{Sun+2025_aspire} are estimated using SED fitting codes, including information from the FIR part of the spectrum in the fit. For all the datasets, stellar masses were estimated from SED fitting codes, and all IMFs were converted to a Chabrier IMF for consistency. To guide the eye, we show a linear relation with a slope of one, generally predicted by simulations and theoretical models, and an arbitrary normalization to highlight the flattening present in the observational data in the low stellar mass range ($\mathrm{log(M_\star/M_\odot) \lesssim 8.5}$), above which our sample is $> 90\%$ mass complete (see Appendix \ref{app:completeness}).

Overall, we find good agreement between the trend identified in this work and that from literature observations, despite the caveats arising from the diverse assumptions adopted in SED fitting (e.g., dust attenuation laws and star formation histories). In particular, we find better agreement with JWST-only observations (CEERS+JADES and CONGRESS), whereas our results tend to fall slightly below the SFR predictions from ALPINE, which include ALMA observations and primarily probe the high-mass end of the relation, where our statistics are limited. The discrepancy with ALPINE data may arise from (i) the incompleteness of the ALPINE sample for stellar masses $\mathrm{M_\star < 10 ^ 9 \; M_\odot}$; (ii) potential underestimation of dust content in massive galaxies within our sample. We have few sources covering the mass range probed by the ASPIRE survey, which has both JWST/NIRCam WFSS in the F356W (along with imaging in F115W, F200W and F356W) and ALMA 1.2 mm continuum observations. However, when we consider the CONGRESS survey - which, like ALT, lacks information from the FIR part of the spectrum and relies solely on JWST/NIRCam grism and imaging, but extends to higher stellar masses - we find that it predicts systematically lower star formation rates. This discrepancy may be due to an underestimation of dust attenuation in CONGRESS, which could suggest a similar effect in ALT (see Sect. \ref{sec:test_dustatt}).

The comparison with literature data highlights the high number statistics of ALT galaxies at the low-mass end of the SFR-M$_\star$ relation: 54\% of the robust H$\alpha$ sample has $\rm log(M_\star/M_\odot) < 8$. By adequately accounting for the selection bias that affects our dataset, we allow for a population-level analysis at $4 < z < 5$ based on a spectroscopic sample that covers the region of the SFR-M$_\star$ plane below $\mathrm{log(M_\star/M_\odot) < 7.5}$. This part of the plane has remained largely unexplored to date; however, see \citet{Rinaldi+2022_sbms} where the authors analyzed a Lyman-$\alpha$ selected sample of galaxies in both blank and lensed fields reaching the low mass end of the SFMS.   

\section{Discussion}
\label{sec:discussion}
\subsection{A flat main sequence slope?}
\label{sec:flat_slope}
In order to interpret our results, we perform detailed comparisons to hydrodynamical simulations, mimicking, as much as possible, the techniques adopted for the observational sample. Hydrodynamical simulations report a unitary slope for the SFR-M$_\star$ relation that does not evolve with redshift \citep[e.g.,][]{Pillepich+2019_tng50, DiCesare+2023, Katz+2023_sphinx, Furlong+2015_eagle, McClymont+2025}. 

\begin{figure*}
   \centering
   \includegraphics[scale=0.4]{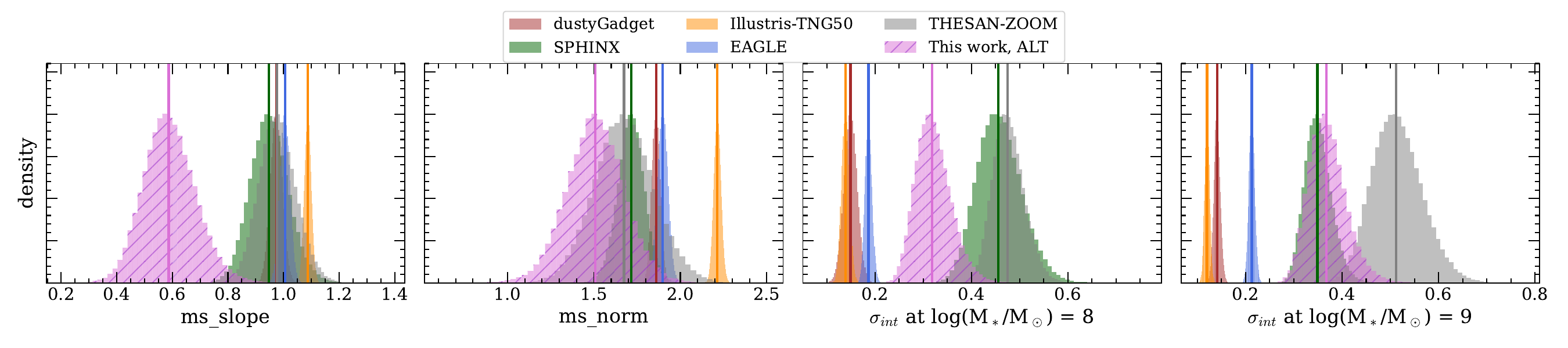}
   \caption{{\bf Comparison between our inferences from observations and hydrodynamical simulations.} Posterior distributions for the slope, normalization and intrinsic scatter of the SFMS. We compare the posteriors based on the ALT robust H$\alpha$ sample (pink, hatched) explored in this work with those from \texttt{dustyGadget} \citep[brown;][]{Graziani+2020_dustyG,DiCesare+2023}, EAGLE \citep[blue;][]{Schaye+2015_eagle,Crain+2015_eagle}, IllustrisTNG-50 \citep[orange;][]{Pillepich+2019_tng50, Nelson+2019_tng50}, SPHINX \citep[green;][]{Rosdahl+2018_sphinx,Rosdahl+2022_sphix,Katz+2023_sphinx} and THESAN-ZOOM \citep[gray;][]{McClymont+2025} simulations.}
   \label{fig:posteriors_sim}
\end{figure*}

Figure \ref{fig:posteriors_sim} shows the posterior distributions obtained by applying the Bayesian model (Sect. \ref{sec:bayesian_model}) to the robust ALT sample as well as to the data from hydrodynamical simulations at $z \sim 4.5$. We consider data from \texttt{dustyGadget} \citep{Graziani+2020_dustyG}, EAGLE \citep{Schaye+2015_eagle, Crain+2015_eagle}, Illustris-TNG50 \citep{Pillepich+2019_tng50}, SPHINX$^{20}$ \citep{Rosdahl+2018_sphinx, Rosdahl+2022_sphix} and THESAN-ZOOM \citep{Kannan+2025_thesanzoom,McClymont+2025} simulations. In particular, the samples used in the Bayesian model consist of stellar masses and star formation rates in each simulation. We apply a SFR cutoff to these simulated galaxy samples, similar to what we do for the observed sample. The specific value of the SFR cutoff varies across simulations, depending on factors such as the number of galaxies remaining after the cut (to ensure a statistically meaningful sample) and  properties of each simulation -- for instance, the SPHINX$^{20}$ data release has a built-in SFR threshold. 

These simulations adopt different numerical strategies and physical assumptions, such as feedback schemes and ISM models, as well as varying mass resolutions and box sizes. For reference, EAGLE has a side length of 100 cMpc, \texttt{dustyGadget}'s side length is 74 cMpc, Illustris-TNG50 span 50 cMpc, and SPHINX covers 20 cMpc, while THESAN-ZOOM is a high-resolution zoom-in simulation targeting galaxies selected from the THESAN parent volume \citep[95.5 cMpc,][]{Kannan+2022-thesan}. Importantly, mass resolution and box size are not independent: larger volumes typically come at the expense of lower mass resolution. While larger simulation boxes allow for better statistics and sampling of massive galaxies, their lower mass resolution prevents them from resolving low-mass systems (e.g. $\rm log(M_\star/M_\odot) < 8$). This trade-off also extends to the ISM modeling. Simulations with coarser resolution (often those with larger volumes) tend to adopt averaged or subgrid ISM models, while higher-resolution simulations can implement more detailed, multi-phase ISM treatments that better capture the structure and physics of the star-forming gas.   

The first two panels of Fig. \ref{fig:posteriors_sim} show that: (i) the best-fit values for the ms\_slope in simulations are $\sim 1$ and are consistent among all of them; (ii) the slope is found to be shallower in our observations; (iii) the ms\_norm at $\rm log(M_\star/M_\odot) = 10.5$ increases with the steepness, as a consequence of the slope-normalization degeneracy. Finally, the last two panels show an increasing intrinsic scatter $\sigma_{int}$ with stellar mass from our observations, whereas simulations show either a scatter that decreases with mass (\texttt{dustyGadget}, SPHINX) or a mass-independent behavior (IllustrisTNG-50, EAGLE). Interestingly, we find that the intrinsic scatter increases with stellar mass in the THESAN-ZOOM simulations. This may be due to the fact that, as zoom-in simulations, THESAN-ZOOM has a lower statistics at high-stellar masses ($\rm log(M_\star/M_\odot) > 10$). Except for SPHINX and THESAN-ZOOM, all simulations show lower values of $\sigma_{int}$ at $\rm log(M_\star/M_\odot)=8$ compared to observational data. This discrepancy among simulations likely arises from the resolution of each simulation and the way in which star formation and feedback mechanisms are implemented within simulation schemes. These physical processes are known to drive stochastic variations in star formation \citep[e.g.,][]{Matthee+2019}, and the inability to resolve such processes in simulations may dampen the variability of star formation histories. This interpretation is supported by the fact that SPHINX and THESAN-ZOOM, which uses the smallest simulation boxes, thus the highest resolution, have the largest $\sigma_{int}$ ($\sim 0.47$ dex) at $\rm log(M_\star/M_\odot) = 8$. Moreover, these two simulations implement the detailed physics of a multiphase ISM \citep{Katz+2023_sphinx, Kannan+2025_thesanzoom}.

It is also worth noting that different simulations trace star formation on different timescales. While for SPHINX and THESAN-ZOOM we consider SFR averaged over 10 Myr, for Illustris-TNG, \texttt{dustyGadget} and EAGLE we use instantaneous SFR estimates. Note that \citet{Pillepich+2019_tng50} found that the scatter in TNG50 is constant regardless of the way SFR is measured (i.e. instantaneous or averaged over 10 or 100 Myr). The median values of the posterior distributions for each parameter are presented in Table \ref{tab:table}. 

Comparison with SFMS trends from both literature observations (Fig. \ref{fig:sfr-mstarALT}) and simulations (Fig. \ref{fig:posteriors_sim}) highlights a shallow slope (ms\_slope $\sim 0.59$) for the H$\alpha$ ALT robust data sample. This slope is inconsistent with predictions from hydrodynamical simulations, and also with theoretical expectations based on the stellar mass function. \citet{Leja+2015} demonstrate that a simple extrapolation of the SFMS trend\footnote{With slope $\alpha(z) = 0.7 - 0.13\,z$} from \citet{Whitaker+2012_sfms} to low-mass galaxies (down to $\rm 10^9 \; M_\odot$) would over-predict the number density of galaxies at $z\sim 2$ by factor 100 (their Fig. 3), see also \citet{Peng+2014}. This growth is the result of the relatively flat slope of the star-forming sequence. The authors argue that the rate of merger interaction cannot compensate for the rapid growth of low-mass galaxies implied by such flat slopes \citep[see also][]{Fu+2024}. If we apply the same logic at $z\sim 4.5$, a slope of $0.59^{+0.10}_{-0.09}$, through the relation between the SFMS and the SMF, would imply a steepening at the low-mass end of the stellar mass function, which is not supported by observations \citep[e.g.,][]{Weibel+2024}.

To investigate the impact of a steeper slope on the posterior distributions of the estimated parameters, we reran the Bayesian model on the robust observational sample, imposing a narrow Gaussian prior on the ms\_slope parameter (mean = 1, $\sigma=$0.02), strongly constraining it to values near unity. In this case, we find an increase in the ms\_norm along with an intrinsic scatter which decreases with stellar mass, more in line with simulations. The intrinsic scatter values are approximately $\sigma_{int} \sim 0.49$ dex at $\mathrm{log(M_\star/M_\odot) = 8}$ and $\sigma_{int} \sim 0.40$ dex at $\mathrm{log(M_\star/M_\odot) = 10}$, as illustrated in Fig. \ref{fig:corner_truncnorm} (purple). Interestingly, this indicates that with our robust sample it is challenging to simultaneously constrain both the slope and the scatter dependence on mass. However, once we fix one parameter (e.g., the slope), we can fit the mass-dependence of the scatter, which decreases with increasing stellar mass.  

\subsection{Testing the effect of varying SFR calibrations}
\label{sec:new_calibrations}

The slope of the SFMS depends on the star formation rate calibrations, which are sensitive to several galaxy properties such as the nature of the massive stellar population (including the role of binary stars), the ISM conditions (i.e. stellar metallicity, dust attenuation), and the assumed IMF \citep[e.g.,][]{Kennicutt1998a, KennicuttEvans2012, Theios+2019}. In this section, we investigate how different SFR calibrations affect the best-fit parameters constrained by our reference model.

Generally, literature calibrations assume a linear relation between the H$\alpha$ luminosity and the SFR, with a constant conversion value (C). Recently, \citet{Kramarenko+2025_sphinx} analyzed SPHINX$^{20}$ simulations at $4.64 \leq z \leq 10$. Using the simulated galaxy data from SPHINX$^{20}$ they selected a sample of star-forming galaxies representative of the H$\alpha$ population observed with JWST at high-redshift. They found that due to the metallicity dependence of the SFR-L$_{\mathrm{H\alpha}}$ relation, the classical calibrations (e.g., \citealt{Kennicutt1998a}), on average, overestimate the SFRs of faint galaxies in SPHINX by $\mathrm{SFR}(\mathrm{H}\alpha)/\mathrm{SFR}_{10} \gtrsim 0.1$ dex. The authors propose two new calibrations: one SFR(H$\alpha$) calibration that depends only on intrinsic H$\alpha$ luminosity (Eq. \ref{eq:SFR(HA)_sphinx}) and the other that additionally depends on the H$\alpha$ equivalent width (Eq. \ref{eq:SFR(HA, EW)_sphinx}), a parameter that is sensitive to stellar metallicity and age.

\begin{equation}
\label{eq:SFR(HA)_sphinx}
    \rm log \left( \frac{SFR}{M_\odot \; yr^{-1}} \right) = 1.06 \; log \left( \frac{L_{H\alpha}}{erg \; s^{-1}} \right) - 43.96
\end{equation}
and 
\begin{align}
\label{eq:SFR(HA, EW)_sphinx}
\rm log \left( \frac{SFR}{M_\odot \; yr^{-1}} \right) & = \rm 0.99 \; log\left(\frac{L_{H\alpha}}{erg \; s^{-1}}\right) \nonumber \\ & \rm  - 0.26 \; log\left( \frac{EW_{0,H\alpha}}{\AA}\right) - 40.34,
\end{align}
\noindent where EW$_{0, \rm H\alpha}$ is the rest-frame equivalent width of the H$\alpha$ emission line.

\begin{figure*}
   \centering
   \includegraphics[scale=0.4]{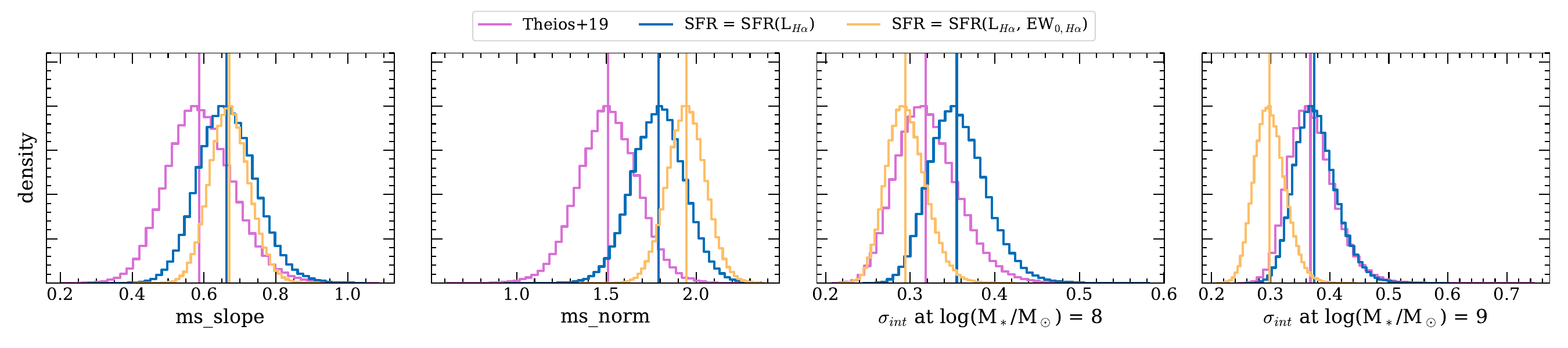}
   \caption{{\bf The impact of new luminosity and EW-dependent SFR(H$\alpha$) calibrations.} Comparison of the posterior distributions for the slope, normalization and intrinsic scatter of the observed SFMS, adopting diverse L$_{\rm H\alpha}$-SFR conversions. In pink are those derived using the calibration from \citet{Theios+2019} (ref.), in blue and yellow are the distributions using calibrations based on SPHINX simulations \citep{Kramarenko+2025_sphinx}. Specifically, in blue we show the posteriors derived from calibrations based on L$_{\rm H\alpha}$ of the galaxies, while in yellow those combining the information from L$_{\rm H\alpha}$ with that from the EW$_{\rm H\alpha}$.}
   \label{fig:posteriors_wSPHINX}
\end{figure*}

We calculate SFRs for the ALT robust H$\alpha$ sample using the alternative calibrations from \citet{Kramarenko+2025_sphinx} and derive new median posteriors with our reference model. Figure \ref{fig:posteriors_wSPHINX} compares the resulting posterior distributions, and Table \ref{tab:table} summarizes the median posterior parameters for slope, normalization, and intrinsic scatter parameters. 

These alternative calibrations produce slightly steeper slopes (up to 14\% increase) and correspondingly higher normalizations. Specifically, using the calibration in Eq. \ref{eq:SFR(HA)_sphinx}, we find ms\_slope $=0.66 \pm 0.08$ and ms\_norm $= 1.79 \pm 0.14 \; \mathrm{M_\odot \; yr^{-1}}$, while with that in Eq. \ref{eq:SFR(HA, EW)_sphinx} we derive ms\_slope $= 0.67 \pm 0.05$ and ms\_norm $= 1.95 \pm 0.11 \; \mathrm{M_\odot \; yr^{-1}}$. The intrinsic scatter shows calibration-independent behavior, remaining approximately constant throughout the explored mass range. 

\subsection{Testing the effect of a non-Gaussian scatter}
\label{sec:non-gaussian}
In our reference model, we adopted truncated normal distributions to model the SFMS scatter accounting for selection effects in our flux-limited sample. However, \citet{McClymont+2025} demonstrate, using THESAN-ZOOM simulations \citep{Kannan+2025_thesanzoom}, that SFRs at fixed mass deviate from log-normal distributions and show a long tail down to zero \citep[see also][]{Feldmann2017, Feldmann2019}. Such a tail, particularly evident at the low-mass end of the relation, may stem from the presence of (mini-) quenched galaxies -- systems in which star formation is halted or strongly suppressed \citep{Dome+2024_mini-q,Looser+2025_miniquenching, Gelli+2025} -- that populate the lower portion of the SFR-M$_\star$ parameter space. 

To investigate the impact of the presence of a low-SFR tail in the SFMS, we modified our reference model to account for a skewed intrinsic scatter, truncated below a threshold value to account for our flux-limited sample selection. We refer to this formulation as the non-Gaussian model. Details can be found in Appendix \ref{app:skewed}. 

Since the values of the skewness parameter (a\_skew) are unknown, we treat it as a free parameter when applying this model to SPHINX and THESAN-ZOOM data. Then, we used a\_skew estimates from simulations as Gaussian priors for this parameter and fit the H$\alpha$ robust observations. We selected data from SPHINX and THESAN-ZOOM as these are the two simulations with higher resolution and smaller boxes, which give insights into the scatter at the low-mass end, where we expect (mini-) quenched galaxies to be.

By fitting SPHINX (THESAN-ZOOM) data with the non-Gaussian model, we derived values of a\_skew = $-2.2 ^{+1.2}_{-1.1}$ (a\_skew = $-4.1 ^{+1.1}_{-1.2}$). These negative values of a\_skew indicate a distribution skewed toward lower SFRs. We adopt these median values and their 1$\sigma$ uncertainties as Gaussian priors in our non-Gaussian model to fit the ALT H$\alpha$ robust sample. With priors from SPHINX data,  we find a slope of ms\_slope = $0.63 ^{+0.14}_{-0.13}$. This result is consistent with the result from THESAN-ZOOM data, indicating that variations in a\_skew have minimal impact on the slope. Moreover, its agreement with the reference Gaussian model shows that introducing skeweness in the SFR does not significantly affect the slope of the SFR-M$_\star$ relation.

\subsection{Testing the effect of dust correction in high-mass galaxies}
\label{sec:test_dustatt}
In this section, we employ a simplified toy model to investigate how a potential underestimation of the dust obscuration at the high-mass end of the SFMS influences the inferred slope and scatter. Figure \ref{fig:dust_corr} presents a comparison between the fraction of obscured H$\alpha$ emission in the ALT robust sample and that of obscured star formation and UV emission reported in the literature \citep{Whitaker+2017, Fudamoto+2020_alpine, Shivaei+2024}. \citet{Whitaker+2017} found a strong dependence of the median $f_{\rm obscured}$ on stellar mass, with little evolution across the redshift range analyzed $0<z<2.5$ \citep[see also][]{Shivaei+2024}. Moreover, they concluded that the transition from mostly unobscured to obscured ($f_{\rm obscured} = 0.5$) star formation occurs at stellar masses $\rm log(M_\star/M_\odot) = 9.4$.

To obtain a fiducial estimate of the potentially underestimated attenuation correction, we implement a mass-dependent dust obscuration correction designed to match the data from \citet{Whitaker+2017} and observations from \citet{Fudamoto+2020_alpine}. This approach is motivated by observational evidence that more massive galaxies tend to have higher dust content and more complex dust geometries \citep[e.g.,][]{Whitaker+2017, Reddy+2025}. However, this method has several caveats. First, we compare the obscured dust fraction inferred from H$\alpha$ emission with that derived from UV emission and UV-based star formation rates. This implicitly assumes that the ratio between stellar and nebular extinction is unity at these redshifts and that the attenuation curves at optical and UV wavelengths are comparable. While these assumptions do not strictly hold, they are adopted here for the sake of illustrating the behavior of our toy model. Second, the model is calibrated to match the obscured fractions reported by \citet{Whitaker+2017}, although \citet{Shivaei+2024} argue that these values may be systematically overestimated. We acknowledge these limitations and emphasize that this exercise is intended solely as a simplified demonstration to explore the qualitative effects of underestimated dust corrections.  

For our toy model, we adopt a sigmoid function to represent a smooth, continuous transition in the strength of the dust correction as a function of stellar mass:

\begin{equation}
    \omega(\mathrm M_\star) = \frac{1}{1 + \exp[s \times (\mathrm M_\star - \mathrm M_{\mathrm center})]}
\end{equation}

\noindent where $\rm M_\star = log(M_\star/M_\odot)$, $\rm M_{center} = log(M_{center}/M_\odot)$ is the mass at which the weight equals 0.5 and $s$ is the steepness parameter controlling the sharpness of the transition. We adopt $\rm M_{center} = 9.5$ and $s=5$, consistent with a rapid transition around the stellar mass value where dust obscuration becomes significantly important. We decrease the H$\alpha$ transmission factor by $f_{\rm eff}$ where $f_{\rm eff} = 1 + (1-\omega) \times (f - 1)$ is the effective correction factor, $(1-\omega)$ is the inverted sigmoid weight that increases with stellar mass, and $f$ is the factor by which the transmission is reduced for the most massive galaxies. We tested correction factors ranging from $f = 1$ (no correction) to $f = 2$ (reducing transmission by half) in steps of 0.2, finding that $f = 2$ is needed to reproduce the obscured faction curve of \citet{Whitaker+2017}, see empty markers in Fig. \ref{fig:dust_corr}. 

We apply these updated dust corrections to the observed H$\alpha$ luminosities -- after correcting for magnification -- to recover intrinsic luminosities, and then compute SFRs using the calibration from \citet{Theios+2019} (Eq. \ref{eq:HatoSFR}). We then fit this updated sample with our Bayesian model and find a slope of ms\_slope = $0.73^{+0.11}_{-0.10}$, together with an increase in normalization, as a consequence of the slope-intercept degeneracy. The intrinsic scatter remains consistent within the uncertainties across the stellar mass range, suggesting only a mild dependence on stellar mass. The median values of the posterior distributions for the parameters, which are consistent with the best-fit values, are summarized in Table \ref{tab:table}.

While the implementation of enhanced dust attenuation corrections for high-mass galaxies - calibrated to match observed obscured star formation fractions from the literature - yields a steeper SFMS slope relative to the ms\_slope obtained from our reference model, we caution that this result is subject to several assumptions. However, we likely underestimate contributions from optically thick star-forming regions, especially in massive galaxies. This bias may explain the comparatively shallower slope we obtain relative to studies incorporating IR emission. Future testing of this hypothesis can be pursued not only through ALMA observations in the IR, but also via JWST/MIRI measurements of Paschen recombination lines, which are less affected by dust attenuation than the commonly used Balmer lines.

\subsection{How to reconcile observations of the SFMS with models?}

In the previous sections, we tested and discussed several scenarios that may help reconcile the tension between our measured (flat) slope of the SFMS and the steeper slopes predicted by theoretical models and hydrodynamical simulations. Such reconciliation is necessary to reliably constrain the mass dependence of the scatter in the SFMS. Specifically, we investigated three key effects: a luminosity-dependent SFR(H$\alpha$) calibration, the presence of a `fat-tail' of galaxies with low SFRs (i.e. non-Gaussian scatter), and the potential underestimation of obscured star formation in our observed sample. Although the first two primarily influence the low-mass end of the SFR-M$_\star$ relation, the third predominantly affects high-mass galaxies. These scenarios impact different aspects of our analysis: some modify the observational measurements (e.g., dust corrections or the conversion of H$\alpha$ luminosity to SFR), while others affect the modeling within our Bayesian framework (e.g. incorporating non-Gaussian scatter).

We found that i) implementing new luminosity-dependent SFR(H$\alpha$) estimates based on SPHINX, with information on EW(H$\alpha$), results in a slightly steeper SFMS slope and an intrinsic scatter of $\sim 0.3$ dex that is independent of stellar mass; ii) adopting a model that allows for non-Gaussian scatter around the SFMS does not significantly affect the slope of the relation. However, the intrinsic scatter shows higher median values as a function of stellar mass, albeit with increased uncertainties; iii) accounting for potentially underestimated dust attenuation in high-mass galaxies ($\rm log(M_\star/M_\odot) > 9$), yields a steeper ms\_slope compared to the reference model and lower median values of the intrinsic scatter with little mass dependence. Among these tests, dust corrections have the strongest impact on the steepness of the SFR-M$_\star$ relation. However, none of these effects alone fully resolves the tension with theoretical models and predictions, suggesting that a combination of all these effects is needed to address this discrepancy.

Follow-up observations of our sample will enable us to observationally test the impact of each proposed effect. In particular, JWST/MIRI, NIRSpec and ALMA observations will allow us to search for characteristic spectral features that relate to the three scenarios presented in this work. JWST/NIRSpec observations of spectral features such as the intensity of the Balmer break and the presence (or not) of emission lines such as H$\beta$ and [\ion{O}{iii}], along with measurements of their equivalent widths, would help us to identify (mini-)quenched galaxies -- which would quantify the extent of non-Gaussian scatter -- and constrain galaxy metallicity. Spectroscopy spanning both rest-frame UV and rest-frame optical wavelengths would provide information on galaxies' SFH which, combined with metallicity measurements, would allow us to test the luminosity-dependent conversion factors found in simulations. Detection of high-SNR Balmer lines would allow for more precise constraints on the Balmer decrement, leading to more accurate estimates of dust attenuation. Combined with JWST/NIRSpec spectroscopy and MIRI coverage of the Paschen recombination lines, ALMA observations of the IR emission from galaxies in our sample would provide complementary constraints on dust-obscured star formation. 

\section{Summary}
\label{sec:summary}
In this work, we studied the relationship between SFR and M$_\star$ in the redshift range $ 3.7 < z < 5.1$, using JWST/NIRCam grism data from the ALT survey. We focused on a robust sample of 316 H$\alpha$ emitters in the region behind the lensing cluster Abell 2744, all with magnification factors $\mu \leq  2.5$. Stellar masses were derived through SED fitting, using all the medium- and broad-band filters available in this field. Notably, 54 \% of the galaxies in our robust sample have $\rm log(M_\star/M_\odot) \leq 8$, enabling us to probe the low-mass end of the SFMS with a statistical sample. Star formation rates were estimated from H$\alpha$ luminosity, adopting the calibration from \citet{Theios+2019}. We further examined how alternative calibrations - based on hydrodynamical simulation predictions - affect the inferred slope, intrinsic scatter, and normalization of the SFMS. Given that our observational sample is biased (i.e. flux-limited survey) toward brighter objects particularly in the low-mass end, we developed a Bayesian model for our analysis of the SFMS which accounts for the selection effect. Our main results are summarized below.

\begin{itemize}
   
\item Using our reference model, we found a slope for the SFMS of ms\_slope = $0.59 ^{+0.10}_{-0.09}$, which is shallower compared to that from literature data \citep[e.g.,][]{Speagle+2014, Popesso+2023} and the one predicted by hydrodynamical simulations \citep[e.g.,][]{Crain+2015_eagle, Pillepich+2019_tng50, Katz+2023_sphinx, DiCesare+2023, McClymont+2025}. The reference model yields intrinsic scatter of 0.3 dex at log($\rm M_\star/M_\odot$) = 8 for the star-forming sequence which increases slightly with stellar mass \citep[see also][]{Cole+2025_ceers}. 

\item By fixing the slope of the SFMS to unity, as predicted by theoretical models and simulations, we found the normalization to be ms\_norm $= 2.09 ^{+0.06} _{-0.07}$ at $\mathrm{log(M_\star/M_\odot)} = 10.5$ and the intrinsic scatter to decrease with increasing stellar mass (see Fig. \ref{fig:corner_truncnorm}). In particular, $\sigma_{int} = 0.46 \pm 0.03$ dex at $\mathrm{log(M_\star/M_\odot)} = 8$ and $\sigma_{int} = 0.41 \pm 0.07$ dex at $\mathrm{log(M_\star/M_\odot)} = 10$ (see Fig. \ref{fig:corner_truncnorm}).

\item We compared the SFRs and stellar masses of the ALT sample with those of other JWST and ALMA surveys at $z\sim4-5$, finding a general good agreement. The high mass end of the ALT data sample falls slightly below the SFR predictions of ALPINE and ASPIRE \citep{Faisst+2020_alpine, Sun+2025_aspire}, which both include ALMA observations. This discrepancy may arise from a potential underestimation of dust content in the massive galaxies within our sample. If this is the case, massive galaxies in our sample would have higher SFR estimates, influencing the slope of the SFMS.

\item We explored alternative SFR calibrations based on SPHINX simulation (Eq. \ref{eq:SFR(HA)_sphinx} and \ref{eq:SFR(HA, EW)_sphinx}, \citealt{Kramarenko+2025_sphinx}). We find that Eq. \ref{eq:SFR(HA, EW)_sphinx}, which includes information on the EW, yields a steeper slope for the SFMS ms\_slope =  $0.67 ^{+0.06} _{-0.05}$, and an intrinsic scatter of $\sim 0.35$ dex that is roughly independent of stellar mass. 

\item Following simulation predictions \citep[e.g.,][]{McClymont+2025}, we investigated the possibility that the scatter around the SFMS is non-Gaussian scatter due to a population of galaxies that are relatively inactive. To explore this, we implemented skewed-normal distributions to model SFRs at fixed stellar masses within our Bayesian framework. We used data from the SPHINX and THESAN simulations to constrain the skewness parameter and consistently find a\_skew $< 0$, indicating a distribution with a tail extending toward lower SFR values. We adopt this simulation-derived skewness as a prior when applying the non-Gaussian model to the ALT robust sample. While this approach yields a slightly steeper slope, ms\_slope $ = 0.63 ^{+0.14}_{-0.13}$, compared to our reference Gaussian model, the difference is modest, indicating that the skewness we introduced in the SFR does not substantially impact the overall SFR-M$_\star$ relation.

\item We tested the impact of dust attenuation in high-mass galaxies on the slope of the SFMS by implementing a mass-dependent dust correction designed to reproduce the obscured fraction reported by \citet{Whitaker+2017} (Fig. \ref{fig:dust_corr}). These updated attenuation values were applied to the observed H$\alpha$ luminosities to derive revised SFRs (Eq. \ref{eq:HatoSFR}). We then fit this updated dust-corrected sample using our reference Bayesian model, which accounts for survey selection function. The resulting slope of the  SFMS, ms\_slope = $0.73^{+0.11}_{-0.10}$, is steeper than that obtained from the original ALT H$\alpha$ robust sample.
\end{itemize}

Although current data do not allow simultaneous constraints on both the slope and the mass-dependent nature of the scatter of the SFMS, our analysis shows that meaningful constraints can be drawn when one of the two is fixed to a known value from independent data. Our analysis underscore the importance of accounting for selection effects that shape observed galaxy samples. Additionally, we propose several testable hypothesis -- such as luminosity-dependent SFR calibrations, the existence of a population of low-mass mini-quenched galaxies, and a systematic underestimation of dust attenuation in high-mass systems -- which are crucial for robustly characterizing the mass dependence of the scatter and the (burstiness of) star formation histories in high-redshift galaxies. 

\begin{acknowledgements}   
    We thank the anonymous referee for the insightful comments that helped improving the manuscript.

     We thank Romain. A. Meyer for valuable discussion, Pierluigi Rinaldi for his help with data handling and Luca Graziani and William McClymont for providing the dustyGadget and THESAN-ZOOM data, respectively. 
     
     Funded by the European Union (ERC, AGENTS,  101076224). Views and opinions expressed are however those of the author(s) only and do not necessarily reflect those of the European Union or the European Research Council. Neither the European Union nor the granting authority can be held responsible for them. 
     
     This work is based on observations made with the NASA/ESA/CSA James Webb Space Telescope. The data were obtained from the Mikulski Archive for Space Telescopes at the Space Telescope Science Institute, which is operated by the Association of Universities for Research in Astronomy, Inc., under NASA contract NAS 5-03127 for JWST. These observations are associated with program \# 3516. We acknowledge funding from {\it JWST} program GO-3516. \\

    Software used in developing this work includes: matplotlib \citep{Hunter2007}, numpy \citep{Oliphant2007}, scipy \citep{Virtanen2020}, TOPCAT \citep{topcat2005}, and Astropy \citep{astropy2013}.

\end{acknowledgements}

\bibliographystyle{aa} 
\bibliography{SFMS} 

\begin{appendix}
\section{Implications for the mass-completeness of the sample}
\label{app:completeness}
\begin{figure}
   \centering
   \includegraphics[scale=0.5]{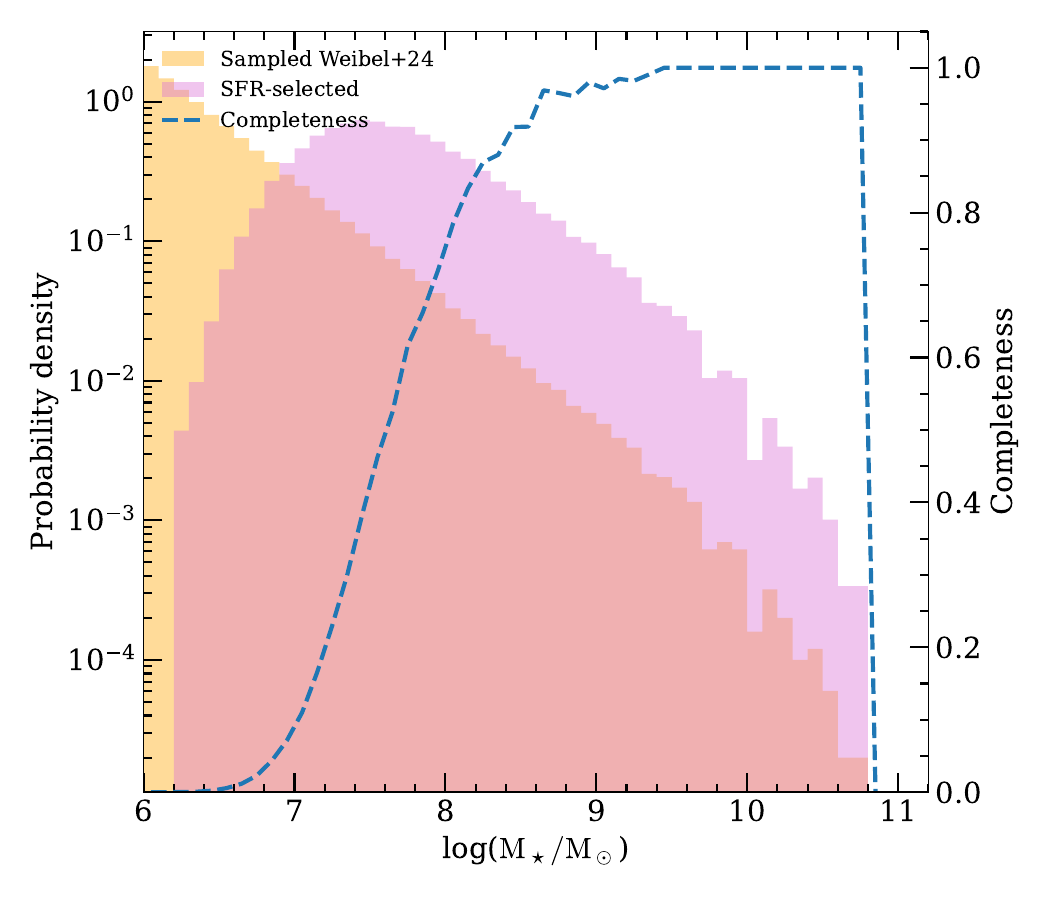}
   \caption{Sampled stellar mass function from \citet{Weibel+2024} (orange), together with the SFR-selected mass function (pink). The completeness trend, derived as the ratio between these two, is shown as dashed blue lines.}
   \label{fig:completeness}
\end{figure}

The mass completeness of a H$\alpha$-limited galaxy sample depends on the relation between H$\alpha$ luminosity and stellar mass. Therefore, from the output of our reference model we can derive the stellar mass completeness of our sample. We computed the mass completeness of our observational sample using the best-fit parameter estimates from the reference model in conjunction with the mass function from \cite{Weibel+2024} (their Table 4). Our approach involved sampling the mass function at $z=5$ across the stellar mass range $\rm log(M_\star/M_\odot) = [6,11]$ with N = 500000 realizations. We then combined these mass values with the best-fit parameters from our reference model (Table \ref{tab:table}) to construct a SFMS. After applying the same SFR cut present in the observations, we generated a new stellar mass function for this filtered sample. The mass completeness of our observational sample was estimated as the ratio between this SFR-selected mass function and the original mass function from \citet{Weibel+2024}, shown in Fig. \ref{fig:completeness}. 

\section{Truncated skewed distribution}
\label{app:skewed}

\begin{figure}[]
   \centering
   \includegraphics[scale=0.45]{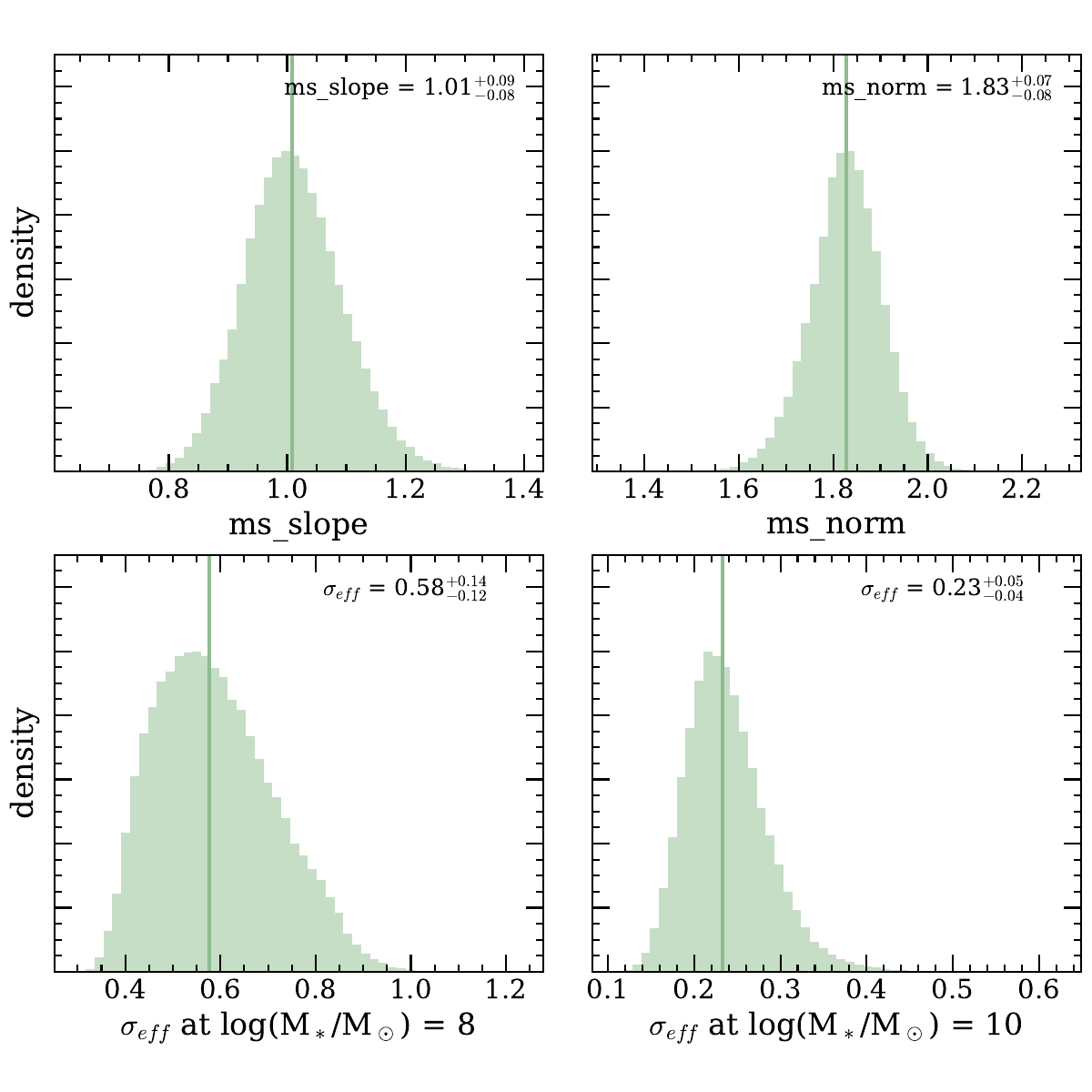}
   \caption{Posterior distributions for the slope, normalization and effective scatter of the SFMS from SPHINX simulation at $z=4.64$.}
   \label{fig:posteriors_sphinx_skew}
\end{figure}

Here we describe the skewed normal distribution implemented in the Bayesian model described in Sect. \ref{sec:bayesian_model}. In particular, the fitted relation in the log-log space is:

\begin{equation}
\label{eq:sfms_skew}
    {\rm log\left( \frac{SFR}{M_\odot \; yr^{-1}}\right)} = \alpha \times {\rm log\left(\frac{M_\star}{10^{10.5} \; M_\odot}\right)} + \beta + \mathcal{SN}(0 \, , \, \sigma_{int} ^2 \, , \, \rm a\_skew)
\end{equation}

\noindent with $\alpha$ = ms\_slope and $\beta$ = ms\_norm, the intercept at $\rm  log(M_\star/M_\odot) = 10.5$. We use the notation $\mathcal{SN}(0 \, , \, \sigma_{int} ^2, \rm a\_skew)$ to indicate a skewed normal distribution with zero mean and variance $\sigma_{eff}^2$, where $\sigma_{eff} = \sigma_{int} \sqrt{ \left[ 1 - 2\delta^2/\pi \right]}$ and $\delta = \rm a\_skew / \sqrt{1 + \rm a\_skew^2}$. The functional form of the intrinsic scatter is the same as in Eq. \ref{eq:int_scatter} and a\_skew is the skewness parameter. Each skewed normal is truncated to mimic the selection effect resulting from flux-limited surveys, resulting in the selection-corrected probability $p$.
We use the log-likelihood function as the objective function for MCMC parameter exploration, identifying the combinations of parameters (ms\_slope, ms\_norm, $\; a, \; b$, a\_skew) that best describe the observed sample by maximizing likelihood.

We apply this model to the full SPHINX catalog at $z = 4.64$ after imposing a cut at SFR $= 0.6 \mathrm{M_\star/M_\odot}$. We multiply our likelihood function $\mathcal{L}$ by a uniform prior probability distribution, with each parameter restricted to the range: $\rm ms\_slope =[0.2,1.8]$, $\rm ms\_norm =[0.01,2.4]$, $a=[-0.1,0.3]$, $b=[0.1,1.3]$ and a\_skew $= [-5,1]$. We use MCMC sampling to explore the parameter space and determine the posterior distribution for each free parameter.  

Figure \ref{fig:posteriors_sphinx_skew} shows the posterior distributions for the slope, normalization and $\sigma_{eff}$ at two stellar masses. The slope and normalization are consistent with those found using truncated normal distributions (see Table \ref{tab:table}). The skewness parameter - which enters $\sigma_{eff}$ via the posterior distribution of a\_skew, $a$ and $b$ - assumes the negative value a\_skew = = $-2.2 ^{+1.2}_{-1.1}$), suggesting a tail towards lower SFRs as predicted from simulations \citep{McClymont+2025}. We also apply this truncated skewed model to the THESAN-ZOOM data, recovering a\_skew $= -4.1 ^{+1.1}_{-1.2}$ and estimates for the slope and normalization consistent with those of Table \ref{tab:table}. 

\end{appendix}
\end{document}